\documentclass[preprint,showkeys,unsortedaddress,superscriptaddress, amsfonts,showpacs,amsmath,amssymb]{revtex4}

\usepackage{euscript,graphicx,amssymb,subfigure}

\usepackage{amsfonts}
\usepackage{amsmath}
\usepackage{amssymb}
\usepackage{graphicx}
\usepackage{euscript}
\usepackage{color}
\usepackage{subfigure}

\newcommand{\goodgap}{\hspace{\subfigtopskip} \hspace{\subfigbottomskip}}

\begin{document}
\title{ Observational constraint in FRW cosmology with a nonminimal scalar field-matter coupling  }

\author{H. Farajollahi$^{1,2}$}
\email{hosseinf@guilan.ac.ir}

\author{A. Salehi$^1$}
\affiliation{$^1$Department of Physics, University of Guilan, Rasht, Iran}
\affiliation{$^2$ School of Physics, University of New South Wales, Sydney, NSW, 2052, Australia}

\date{\today}

\begin{abstract}
 \noindent \hspace{0.35cm}
 In this paper within the scope
of FRW cosmology for $k=0, \pm 1$ we study the dynamics of the universe for a cosmological model with a scalar field nonminimally coupled to matter. By best-fitting the model parameters with the observational data for the direct interaction between the dark sectors in the model we obtain observational constraints on cosmological parameters. The result shows that with the best fitted model parameters, only in flat universe, the phantom crossing occurs twice in the past and once in the future, whereas no crossing occurs for open and closed models of the universe.

\end{abstract}


\maketitle

\section{introduction}

Recently, the observations of high redshift type Ia
supernovae and the surveys of clusters of galaxies \cite{Reiss}--\cite{Pope} reveal the universe accelerating expansion
and that the density
of matter is very much less than the critical density. Also the
observations of Cosmic Microwave Background (CMB)
anisotropies indicate that the universe is flat and the total energy
density is very close to the critical one \cite{Spergel}.

The above observational data properly complete each other and point out
that the dark
energy (DE) is the dominant component of the present universe which occupies about $\%73$ of the energy of
our universe, while dark matter (DM) occupies $\%23$, and the usual
baryonic matter about $\%4$. There are prominent candidates for DE such as the cosmological
constant \cite{Sahni, Weinberg}, a dynamically evolving scalar field ( like quintessence)
\cite{Caldwell, Zlatev} or phantom (field with negative energy) \cite{Caldwell2} that explain the cosmic accelerating expansion. Meanwhile, the accelerating
expansion of universe can also be described through
modified gravity \cite{Zhu},  brane cosmology and so on \cite{Zhu1}--\cite{Easson}.

To explain the early and late time acceleration of the
universe. it is most often the case that such fields interact with matter; directly due to a matter Lagrangian
coupling, indirectly through a coupling to the Ricci scalar or as the result of quantum loop corrections \cite{Damouri}--\cite{Biswass}. If the
scalar field self-interactions are negligible, then the experimental bounds on such a field are very strong; requiring it to either
couple to matter much more weakly than gravity does, or to be very heavy \cite{Uzan}--\cite{Damourm}. In some models, coupling of the scalar field to matter should be tuned to
extremely too small values in order not to be conflict with the Equivalence Principal \cite{nojiri}. In
the proposed  models, the scalar field couples to matter with gravitational strength, in harmony with general expectations from
string theory \cite{Khoury}--\cite{Khourym}. Then, the scalar field is permitted
to couple to matter much more strongly than gravity does, and yet still satisfies the current experimental and observational
constraints. The cosmological value of such a field evolves over Hubble time-scales and could potentially cause the late-time acceleration of our Universe \cite{Brax2}. The crucial feature that these models possess are that the mass of the scalar field depends on the
local background matter density. While the idea of a density-dependent mass term is not new \cite{Wett}--\cite{Mot}, the
work presented in \cite{Khourym, Brax2} is novel in that the scalar field can couple directly to matter with gravitational strength. Since the scalar field mimics the background (radiation/matter) matter field, subdominant for most of the evolution history except at late times  when it becomes dominant, it may be regarded as a cosmological tracker field \cite{farajollahi-jcap}.

In this paper, we study the detailed evolution of a cosmological model that an scalar field nonminimally coupled to the matter lagrangian. We assume FRW cosmological model with open, closed and open background geometries.
In Sec. 2, we present the model in terms of dimensionless dynamical variables. In Sec. 3, by best-fitting the model with the most recent observational data, the Type Ia supernovea (SNe Ia)for distance modulus and applying $\chi^2$ method, new constraints on the model parameters are revealed. Sec. 4 devoted to the cosmological test of the model by examining the behavior of the effective EoS parameter and also reconstructing the model evolution from the solution. The interaction model is presented in Sec. 5. We summarize our results in section 6.

\section{The Model}

In this section we consider an action of gravity in the presence of a scalar field nonminimally coupled to the matter lagrangian,
\begin{eqnarray}\label{action}
S=\int[\frac{R}{16\pi
G}-\frac{1}{2}g^{\mu\nu}\nabla_\mu\phi\nabla_\nu\phi+V(\phi)+f(\phi)L_{m}]\sqrt{-g}dx^{4},
\end{eqnarray}
where $R$ is Ricci scalar, $G$ is the newtonian constant gravity
and $\phi(x^\mu)$ is the scalar field with a potential
$V(\phi)$. Unlike the usual Einstein-Hilbert action, the matter
Lagrangian $L_{m}$ is modified as $f(\phi)L_{m}$, where $f(\phi)$ is
an analytic function of the scalar field. This last term in Lagrangian brings about the nonminimal
interaction between the matter and scalar field.
The variation of action (\ref{action})  with respect to the metric tensor components in FRW  cosmology
yields the field equations,
\begin{eqnarray}\label{fried1}
3(H^{2}+\frac{k}{a^2})=\rho_{m}f+\frac{1}{2}\dot{\phi}^{2}+V(\phi),
\end{eqnarray}
\begin{eqnarray}\label{fried2}
2\dot{H}+3H^2+\frac{k}{a^2}=-\frac{1}{2}\dot{\phi}^{2}+V(\phi),
\end{eqnarray}
where we put  $8\pi G=c=\hbar=1$ and assume a perfect fluid with $p_{m}=\gamma\rho_{m}$. In the above equations, for flat, closed and open models of the universe, $k=0, \pm 1$, respectively. We also assumed that he energy density $\rho_{m}$ stands for the contribution
from the cold dark matter to the energy density, $\gamma=0$. Also
variation of the action (\ref{action}) with respect to the scalar field  $\phi$ provides the wave
equation for the scalar field as
\begin{eqnarray}\label{phiequation}
\ddot{\phi}+3H\dot{\phi}=-V^{'}-\rho_{m}f^{'},
\end{eqnarray}
where prime indicated differentiation with respect to $\phi$. One can easily find that both components -the scalar field $\phi$ and the coupling $f(\phi)$ with CDM, $\rho_m f$- do not conserve separately but that they interact through a term $Q$ according to
\begin{eqnarray}\label{roef1}
\dot{(\rho_{m}f)}+3H\rho_{m}f=Q,
\end{eqnarray}
\begin{eqnarray}\label{pef1}
\dot{\rho_{\phi}}+3H(1+\omega_{\phi})\rho_{\phi}=-Q,
\end{eqnarray}
where $Q =\rho_m f'\dot{\phi}$ is the energy exchange term derived from the model. We also define the ratio between the energy densities as $r\equiv \frac{\rho_{m}f}{\rho_{\phi}}$.

In the following we present the structure of the dynamical system by introducing the dimensionless variables:
\begin{eqnarray}\label{defin}
 \Omega_{\dot{\phi}}=\frac{\dot{\phi}^{2}}{6 H^{2}},\ \ \Omega_{mf}={\frac{\rho_{m}f}{3 H^{2}}} ,\ \ \Omega_{V}=\frac{V}{3H^{2}},\ \ \Omega_{k}=\frac{k}{3a^{2}H^{2}},
\end{eqnarray}
where "$mf$" means coupling of the scalar function $f$ with matter field.

We consider that both scalar function and scalar potential behave exponentially as $f(\phi)=f_{0}\exp{(\alpha\phi)}$ and $V(\phi)=V_{0}\exp{(\beta \phi)}$ where $\alpha$ and $\beta$ are  dimensionless constants characterizing the slope of potential $V(\phi)$ and $f(\phi)$. The cosmological models with such exponential functions have been known lead to interesting physics
in a variety of context, ranging from existence of accelerated expansions \cite{Halliwell} to cosmological
scaling solutions \cite{Ratra}--\cite{Yokoyama}. In particular, the exponential forms of $f(\phi)$ and $V(\phi)$ are motivated by a variety of cosmological models \cite{chameleon} and also from stability considerations \cite{stability}. By using equations (\ref{fried1})-(\ref{phiequation}), the evolution equations of these variables become,
\begin{eqnarray}
\frac{d\Omega_{\dot{\phi}}}{dN}&=&-2\Omega_{\dot{\phi}}\frac{\dot{H}}{H^{2}}
-\frac{\sqrt{6\Omega_{\dot{\phi}}}}{2}\frac{\ddot{\phi}}{H^{2}}
,\label{x1}\\
\frac{d\Omega_{mf}}{dN}&=&-2\Omega_{mf}\frac{\dot{H}}{H^{2}}-3\Omega_{mf}
+\alpha\Omega_{mf}\sqrt{6\Omega_{\dot{\phi}}},\label{y1}\\
\frac{d\Omega_{V}}{dN}&=&\Omega_{V}[\sqrt{6}\beta\Omega_{\dot{\phi}}^{\frac{1}{2}}+\Omega_{mf}
+6\Omega_{mf}^{2}],\label{z1}\\
\frac{d\Omega_{k}}{dN}&=&-2\Omega_{k}-2\Omega_{k}\frac{\dot{H}}{H^{2}},\label{k1}
\end{eqnarray}
where $N \equiv ln (a)$ and
\begin{eqnarray}
\frac{\dot{H}}{H^{2}}&=&\frac{-1}{2}[3\Omega_{mf}+6\Omega_{\dot{\phi}}]+\Omega_{k},\label{hdot}\\
\frac{\ddot{\phi}}{H^{2}}&=&\sqrt{6}\Omega_{\dot{\phi}}^{\frac{1}{2}}+\beta\Omega_{V}
+\alpha\Omega_{mf}.\label{phidot}
\end{eqnarray}
In term of the new dynamical variables, also the Friedmann constraint is
\begin{eqnarray}\label{constraint}
\Omega_{k}+1=\Omega_{mf}+\Omega_{V}+\Omega_{\dot{\phi}}.
\end{eqnarray}
One can write the effective deceleration and EoS parameters for the model and also the EoS parameter for the scalar field in terms of new variables as
 \begin{eqnarray}
q&=&-1-\Omega_{k}+3\Omega_{\dot{\phi}}+\frac{3}{2}\Omega_{mf},\label{q}\\
\omega_{\phi}&=&\frac{\Omega_{\dot{\phi}}-\Omega_{V}}{\Omega_{\dot{\phi}}+\Omega_{V}},\label{phidot}\\
\omega_{eff}&=&\frac{\Omega_{\dot{\phi}}-\Omega_{V}}{\Omega_{mf}+\Omega_{\dot{\phi}}
+\Omega_{V}}\label{eff}.
\end{eqnarray}
By imposing the constraint (\ref{constraint}) the equations (\ref{x1})-(\ref{k1}) reduce to three equations for the dynamical variables $\Omega_{\dot{\phi}}(0)$, $\Omega_{mf}(0)$ and $\Omega_{V}(0)$ .

In the next section we solve the above equations by best fitting the model parameters  $\alpha$, $\beta$ and initial conditions $\Omega_{\dot{\phi}}(0)$, $\Omega_{mf}(0)$, $\Omega_{V}(0)$ and $H(0)$ with the observational data for distance modulus using the $\chi^2$ method. The advantage of simultaneously solving the system of equations and best fitting the model parameters is that the solutions become physically meaningful and observationally favored.

\section{Cosmological constraints}

The difference between the absolute and
apparent luminosity of a distance object is given by, $\mu(z) = 25 + 5\log_{10}d_L(z)$ where the luminosity distance quantity, $d_L(z)$ is given by
\begin{equation}\label{dl}
d_{L}(z)=(1+z)\int_0^z{-\frac{dz'}{H(z')}}.
 \end{equation}
With numerical computation, we solve the system of dynamical equations for $\Omega_{\dot{\phi}}$, $\Omega_{k}$, $\Omega_{mf}$ and $\Omega_{V}$. While best fitting the model parameters and initial conditions with the most recent observational data, the Type Ia supernovea (SNe Ia), in order to accomplish the mission, we need the following two auxiliary equations for the luminosity distance and the hubble parameter
\begin{eqnarray}
 \frac{dH}{dN}=H(-\frac{\dot{H}}{H^{2}}),
 \end{eqnarray}
\begin{eqnarray}
\frac{d(d_{L})}{dN}=-(d_{L}+\frac{e^{-2N}}{H}).
\end{eqnarray}
To best fit the model for the parameters $\alpha$, $\beta$ and the initial conditions $\Omega_{\dot{\phi}}(0)$, $\Omega_{mf}(0)$, $\Omega_{V}(0)$ and $H(0)$ with the observational data, SNe Ia, we employe the $\chi^2$ method. From now on, we also assume that the matter field in the model is cold dark matter (CDM), $\gamma=0$. We constrain the parameters including the initial conditions by minimizing the $\chi^2$ function given as
\begin{widetext}
\begin{equation}\label{chi2}
 \chi^2_{SNe} ( \alpha,\beta; \Omega_{\dot{\phi}}, \Omega_{mf}, \Omega_{V}, H|_0)=\sum_{i=1}^{557}\frac{[\mu_i^{the}(z_i|\alpha,\beta; \Omega_{\dot{\phi}}, \Omega_{mf}, \Omega_{V}, H|_0) - \mu_i^{obs}]^2}{\sigma_i^2},
\end{equation}
\end{widetext}
where the sum is over the SNe Ia data. In relation (\ref{chi2}), $\mu_i^{the}$ and $\mu_i^{obs}$ are the distance modulus parameters obtained from our model and observation, respectively, and $\sigma$ is the estimated error of the $\mu_i^{obs}$. Table I shows the best fitted model parameters and initial conditions in all three cases, $k=0, \pm 1$.\\

\begin{table}[ht]
\caption{Best-fitted model parameters} 
\centering 
\begin{tabular}{c c c c c c c c} 
\hline 
model  &  $\alpha$  &  $\beta$ \ & $\Omega_{\dot{\phi}}(0)$\ & $\Omega_{mf}(0)$\ & $\Omega_{V}(0)$ \ & $H(0)$
\ & $\chi^2_{min}$\\ [2ex] 
\hline 
$k=0$&$1.8$  & $2.2$ \ & $-0.4$\ & $-0.2$\ & $1.6$\ & $0.91$ \ & $548.0229449$ \\
$k=1$ & $0.94$  & $-0.07$ \ & $0.7$\ & $-0.2$\ &$2.05$\ &  $0.916$ \ & $557.5901281$ \\
$k=-1$&$6.5$  & $3.7$ \ & $0.4$\ & $0.5$\ & $2.2$\ & $0.908$ \ & $546.03629$ \\
\hline 
\end{tabular}
\label{table:1} 
\end{table}\

\begin{figure*}
\centering
\subfigure{\includegraphics[width=7cm]{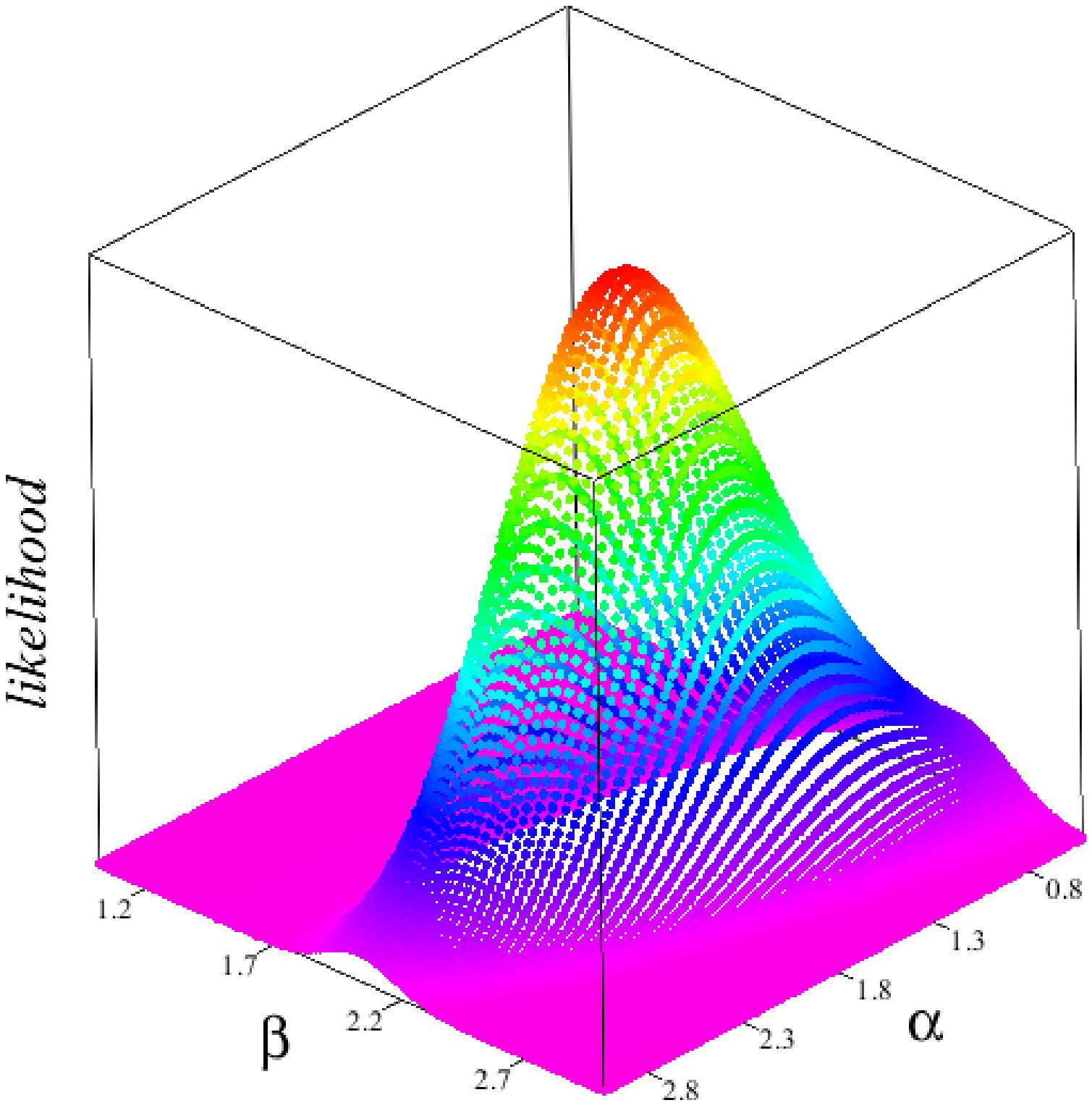}} \goodgap
\subfigure{\includegraphics[width=7cm]{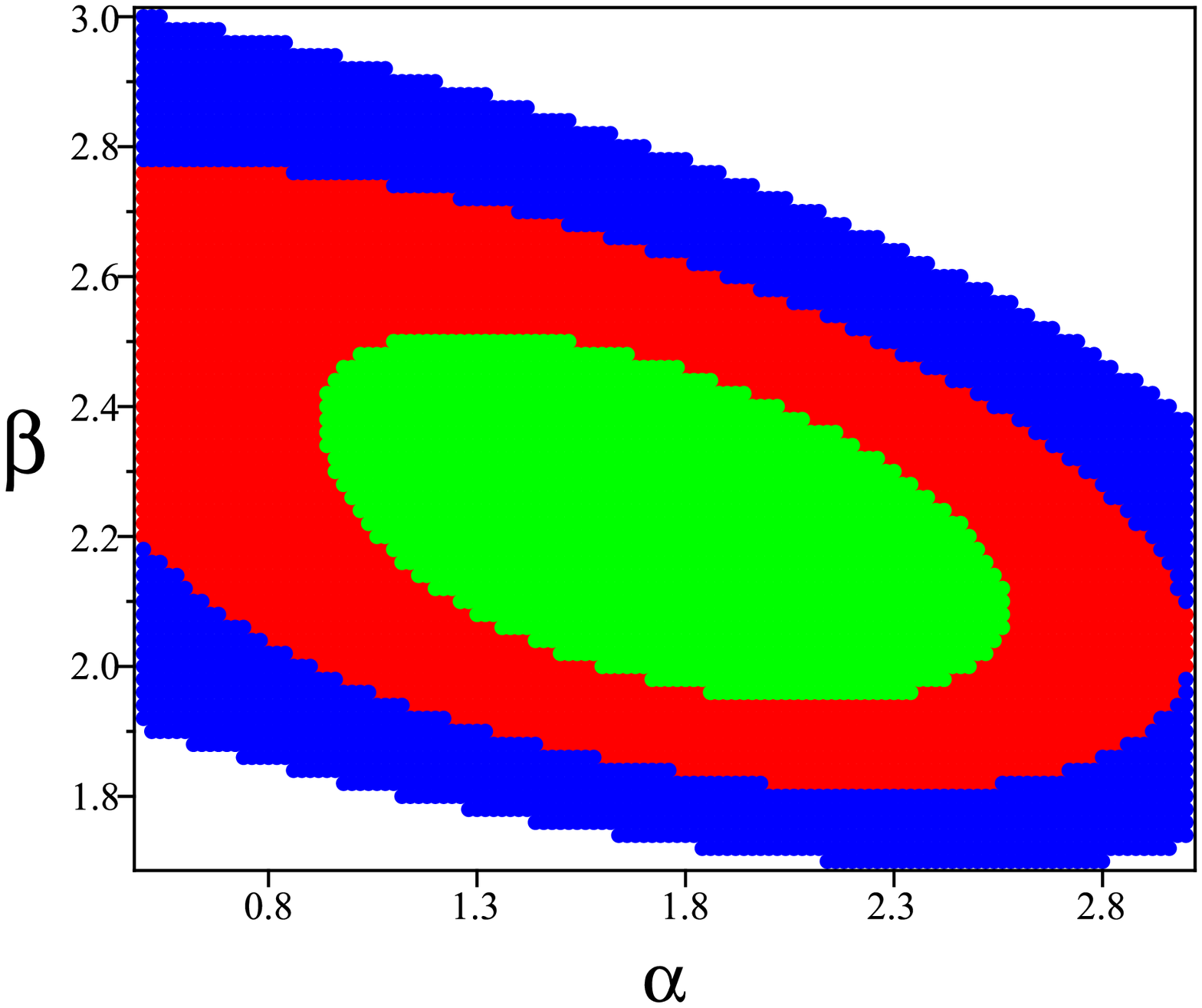}}
\caption{The best-fitted two dimension likelihood and confidence level for $\alpha$, $\beta$ in $k=0$ case}
\label{fig: clplots}
\end{figure*}
\begin{figure*}
\centering
\subfigure{\includegraphics[width=7cm]{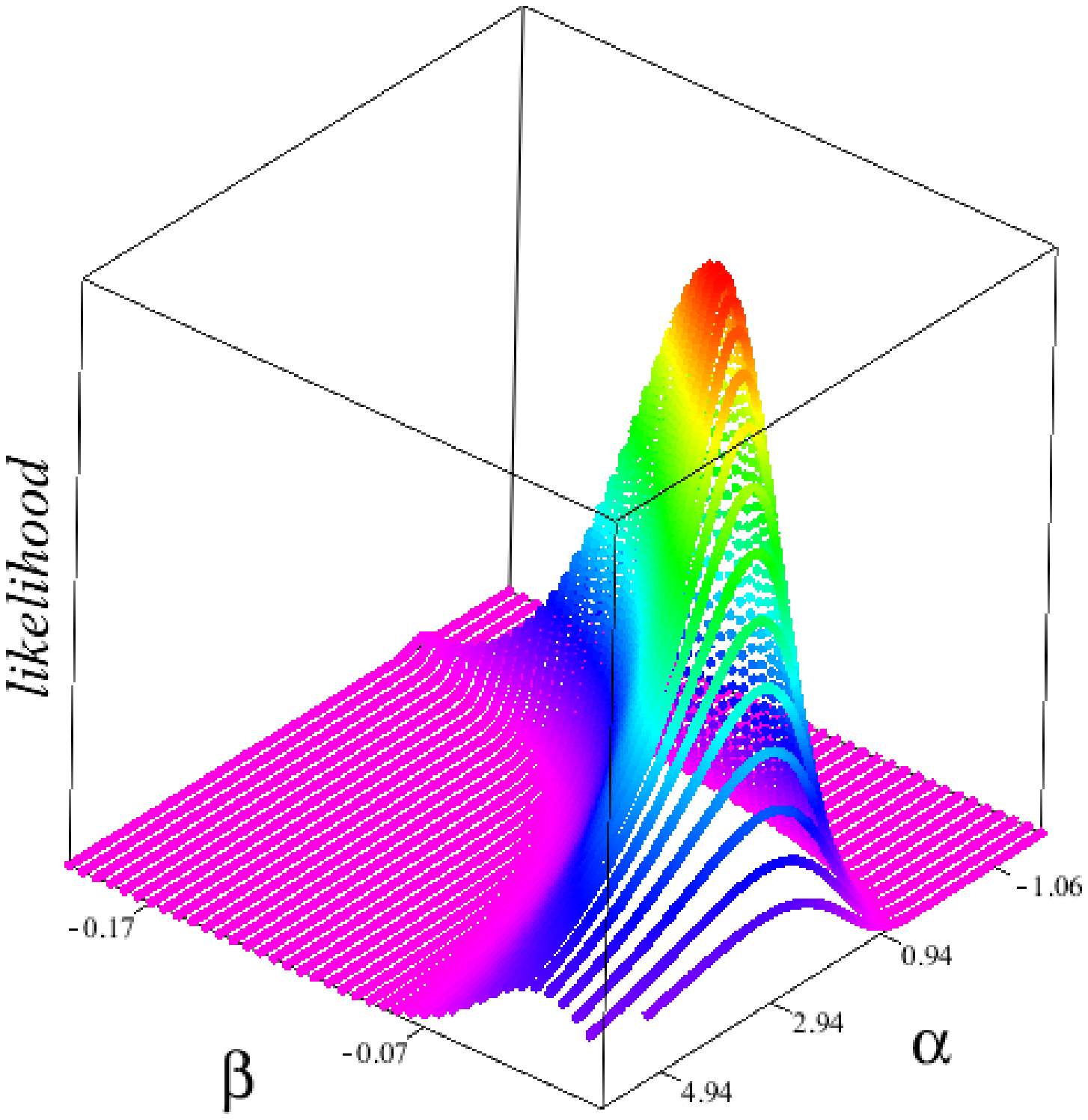}} \goodgap
\subfigure{\includegraphics[width=7cm]{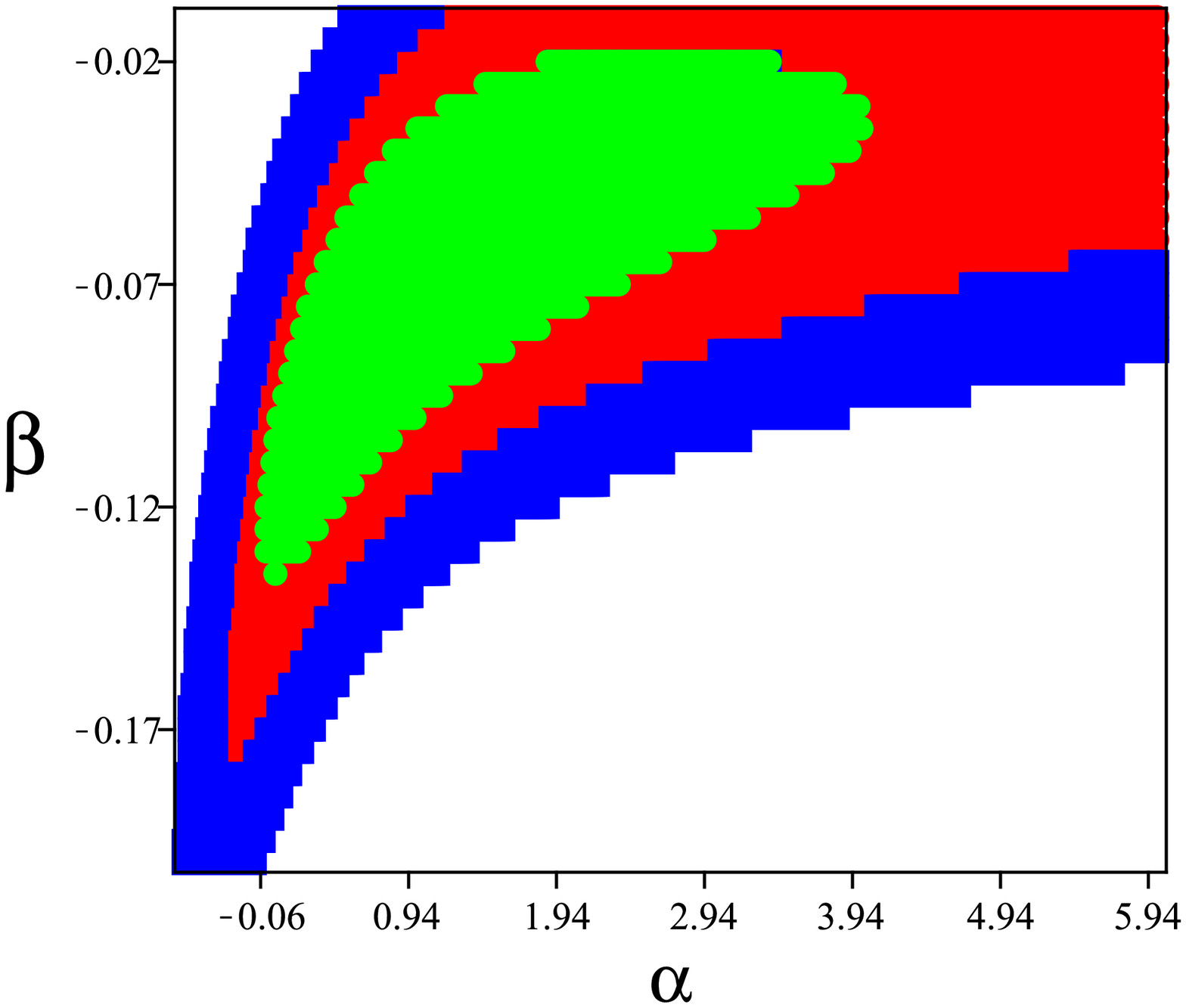}}
\caption{The best-fitted two dimension likelihood and confidence level for $\alpha$, $\beta$ in $k=+1$ case}
\label{fig: clplots}
\end{figure*}
\begin{figure*}
\centering
\subfigure{\includegraphics[width=7cm]{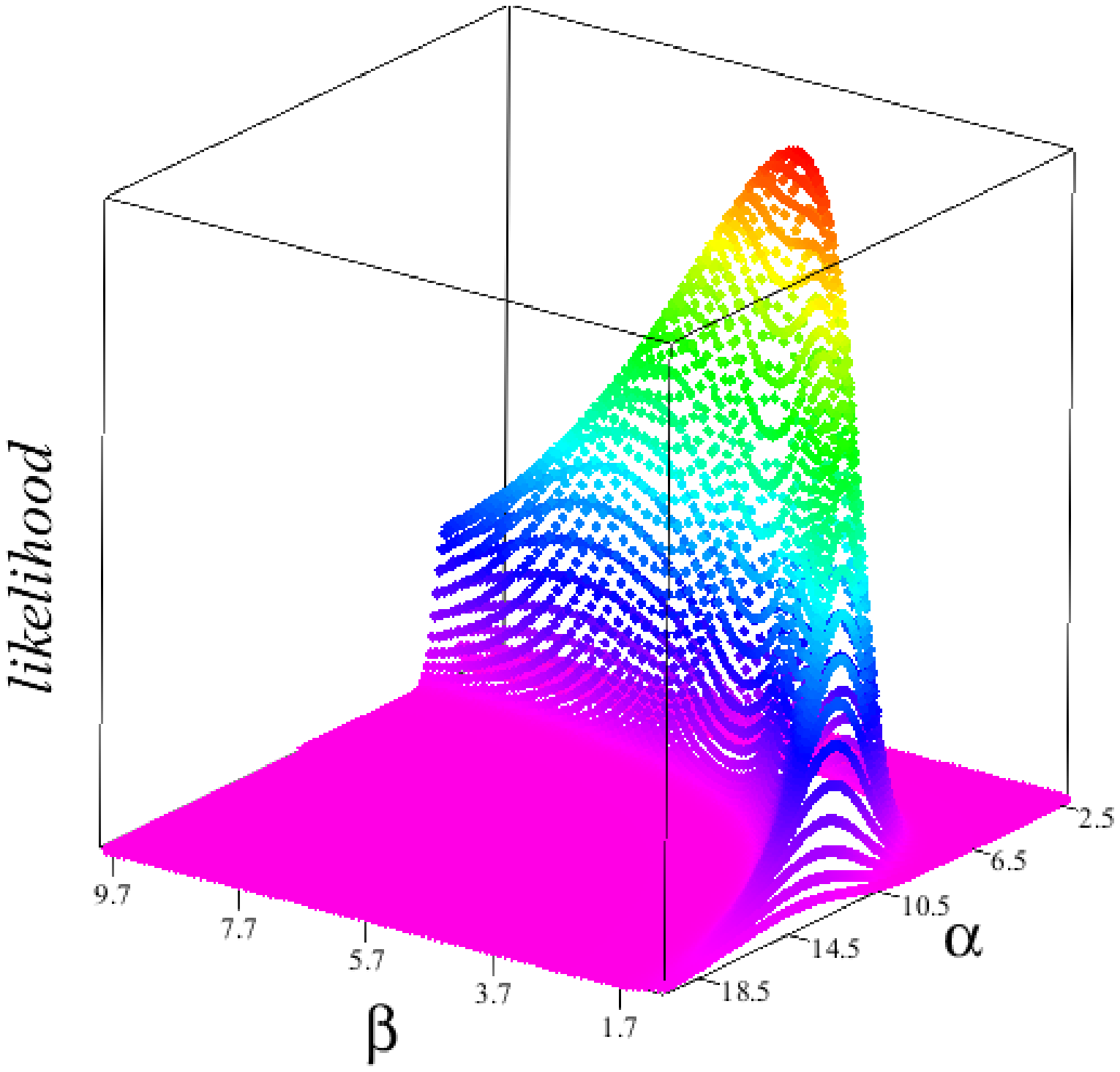}} \goodgap
\subfigure{\includegraphics[width=7cm]{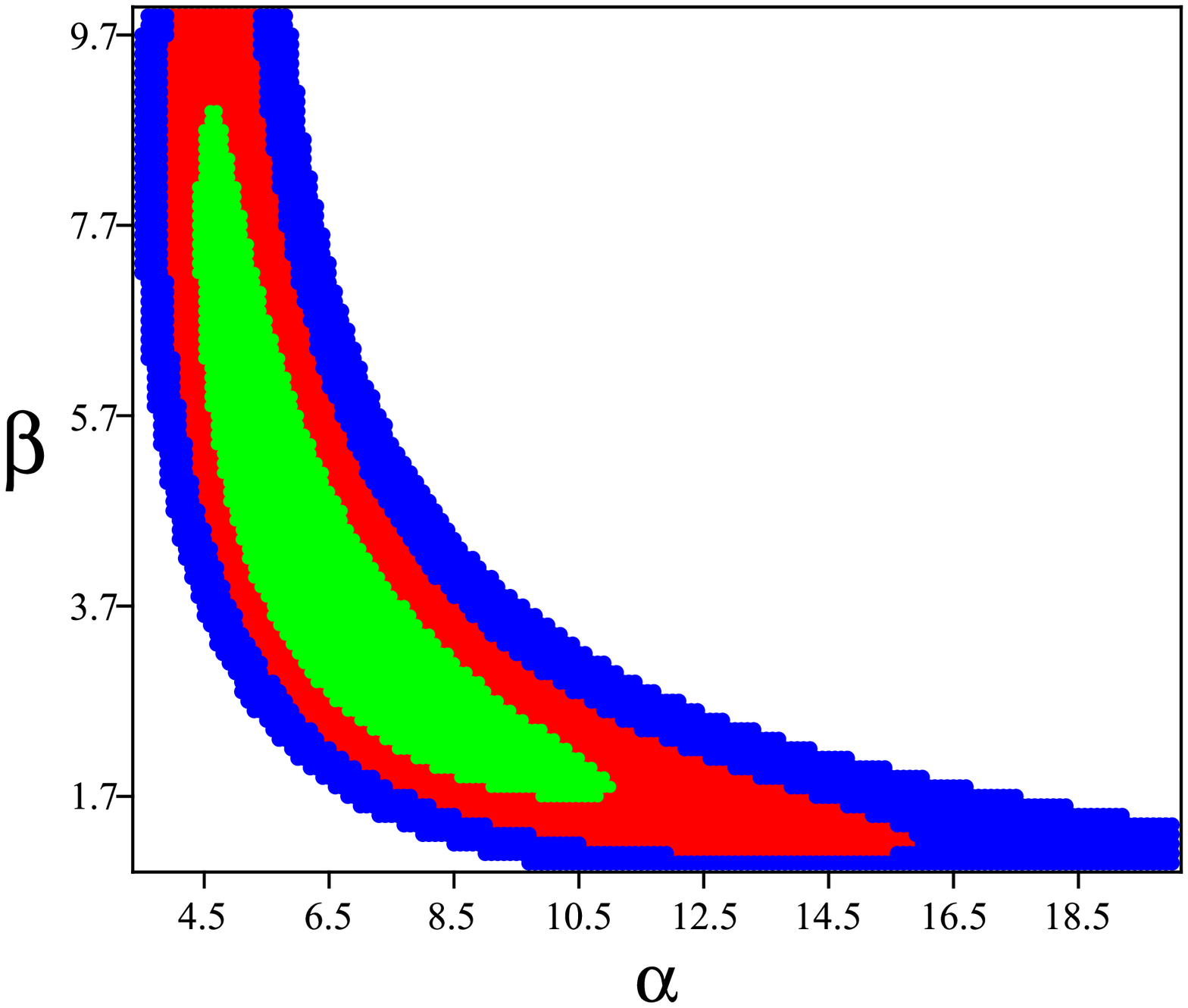}}
\caption{The best-fitted two dimension likelihood and confidence level for $\alpha$, $\beta$ in $k=-1$ case}
\label{fig: clplots}
\end{figure*}\

Figs. 1-3 show the constraints on the parameters $\alpha$, $\beta$ with the initial conditions on $\Omega_{\dot{\phi}}(0)$, $\Omega_{mf}(0)$, $\Omega_{V}(0)$ and $H(0)$  at the $68.3\%$, $95.4\%$ and $99.7\%$ confidence levels. The best fitted 2 dimensional likelihood the model parameters are also shown in figures 1-3.

In Fig. 4, the distance modulus, $\mu(z)$, in our model is fitted with the observational data for the parameters $\alpha$, $\beta$ and initial conditions $\Omega_{\dot{\phi}}(0)$, $\Omega_{mf}(0)$, $\Omega_{V}(0)$ and $H(0)$, using $\chi^2$ method in all three cases $k=0, \pm 1$.\\

\begin{figure*}
\centering
\subfigure{\includegraphics[width=7cm]{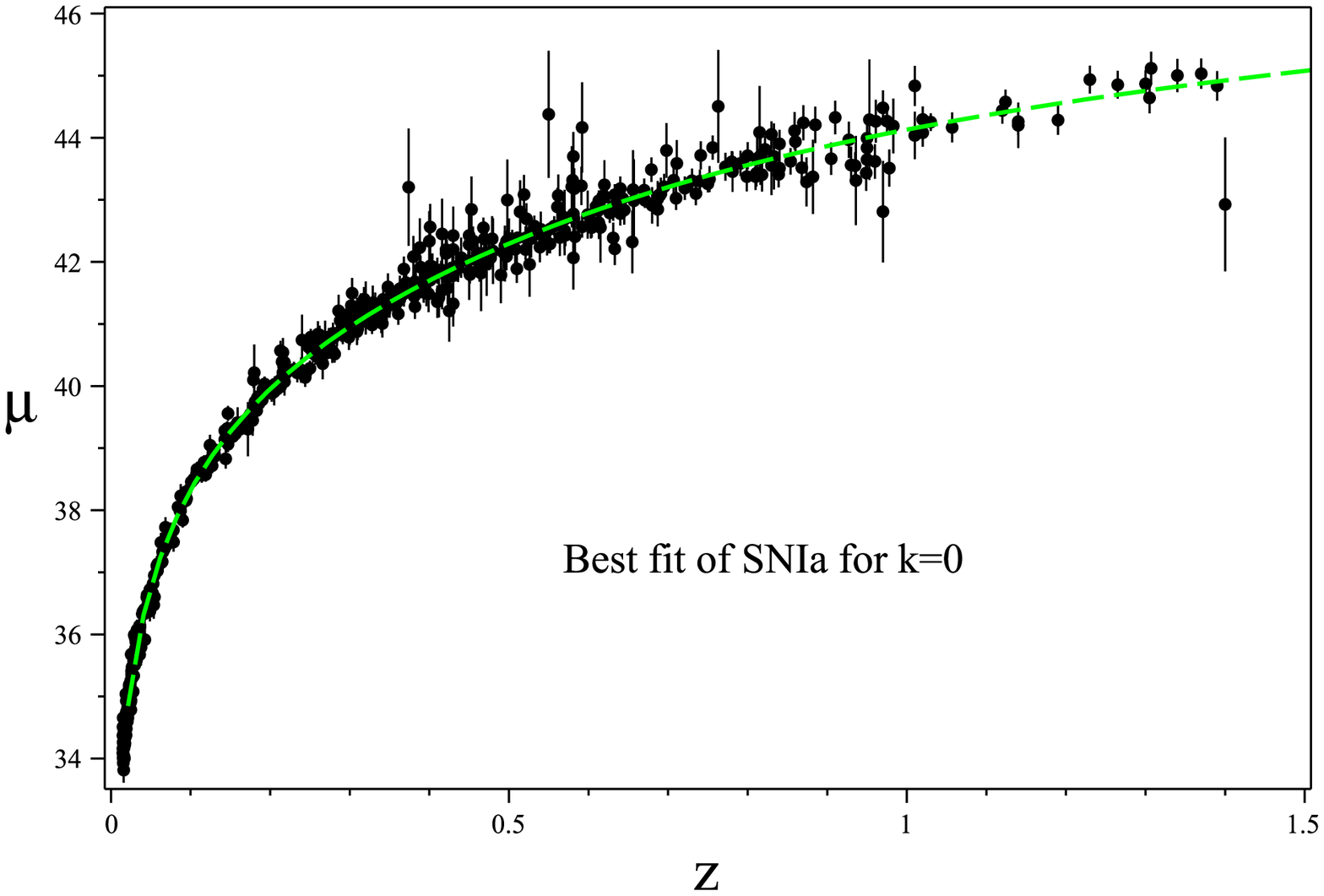}} \goodgap
\subfigure{\includegraphics[width=7cm]{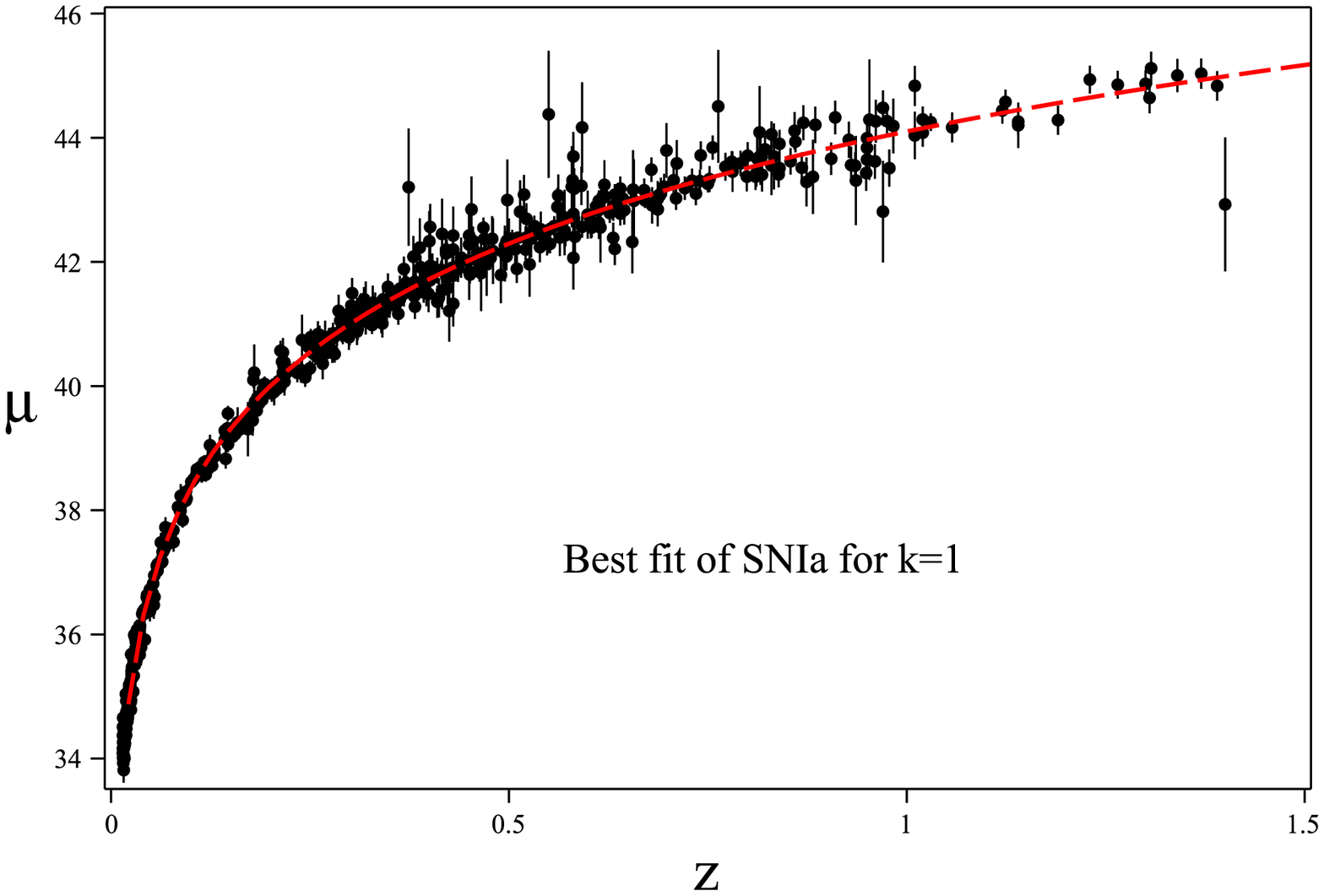}} \goodgap
\subfigure{\includegraphics[width=7cm]{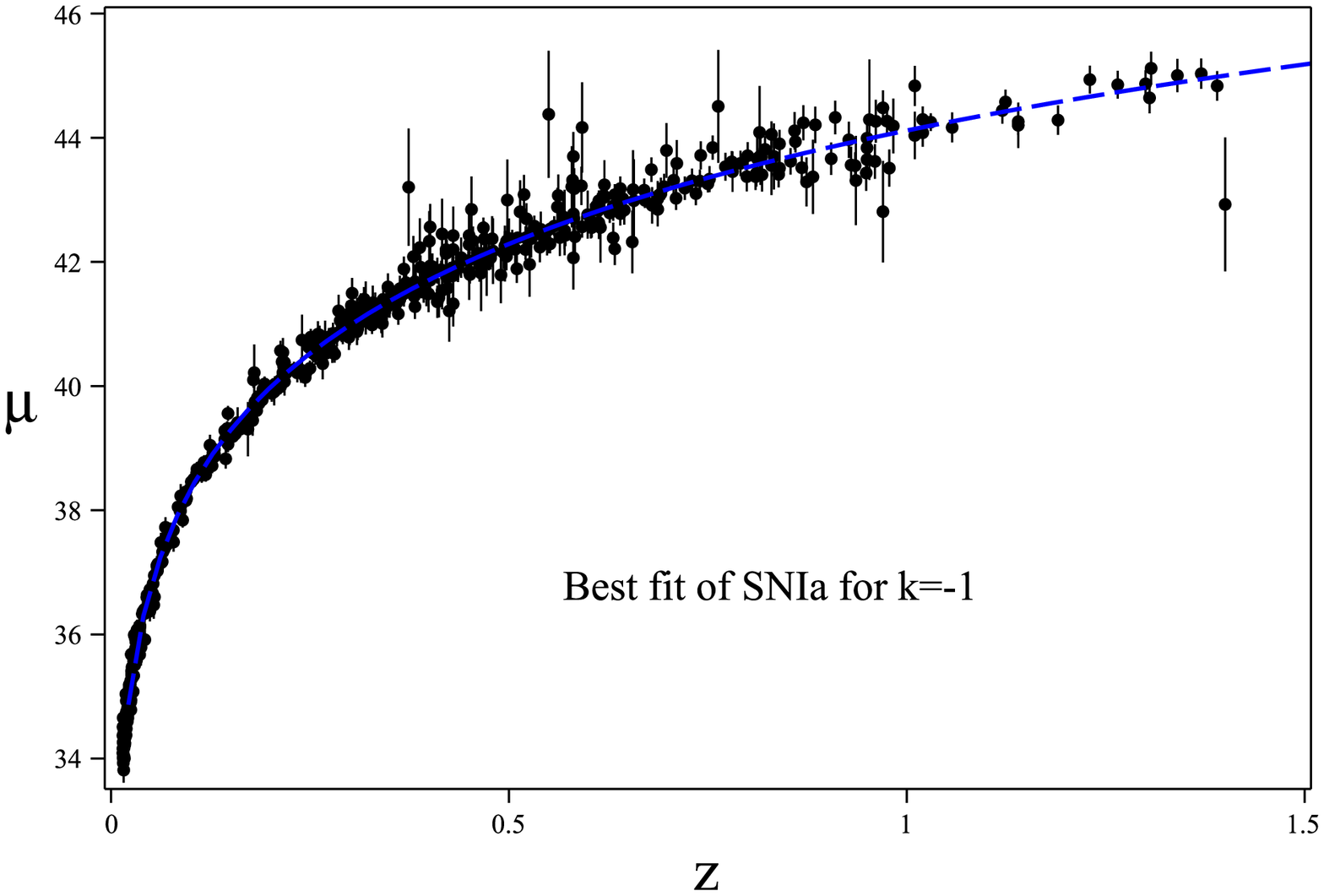}} \goodgap\\
\caption{The best-fitted distance modulus $\mu(z)$ plotted as function of redshift in $k=0, \pm 1$ cases }
\label{fig: clplots}
\end{figure*}

\section{Results and discussion}

In order to realize the behavior of the universe and its dynamics we need to verify the best-fitted model parameters. We have already best fitted our model with the current observational data for the distance modulus. The cosmological parameters analytically and/or numerically have been investigated by many authors for variety of cosmological models. The effective EoS and deceleration parameters, plus scalar field EoS parameter are given by the equations (\ref{q})-(\ref{eff}).

In Fig. 6, these parameters are shown for all three models of the universe, $k=0, \pm 1$.  All the trajectories shown in these graphs are best fitted for the model  parameters $\alpha$ and $\beta$ and also initial conditions. From Fig. 6)left for scalar field EoS parameter one observes that in none of the closed, open and flat models of the universe the phantom crossing occur in the past and future. In $k=0$ case, The EoS parameter for the scalar field, $\omega_{\phi}$, become tangent to the phantom divide line in the near past. For effective EoS parameter, Fig. 6)middle, phantom crossing occurs twice in the past and once in the future in flat model of the universe. The phantom crossing can also can be verified from Fig. 5, for $k=0$ three times whenever $2\Omega_{\phi}=-\Omega_{mf}$ for $f<0$ and for $k=\pm 1$ never occur, as the conditions are not satisfied. The graph also shows in each case of open and closed universe, the behavior of EoS parameters are very similar to each other. The graphs for both EoS parameters in case of open universe also show a singularity in the near past. The deceleration parameter for all three cases exhibits a universe transition from deceleration state to acceleration one in the near past. In case of open universe the graph for the deceleration parameter approaches minus infinity in the near future, a phantom superacceleration that leads to Big Rip singularity.

\begin{figure*}
\centering
\subfigure{\includegraphics[width=7cm]{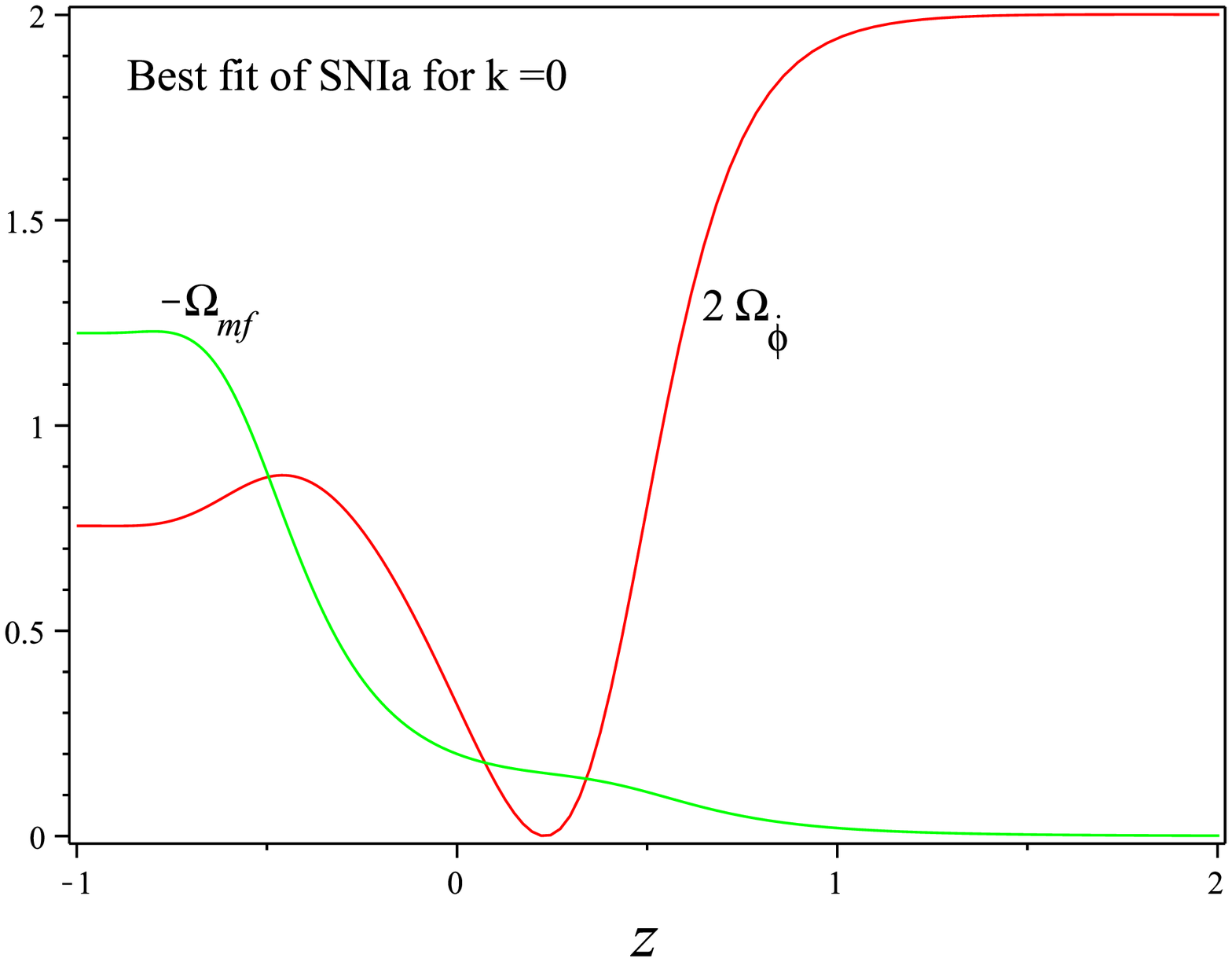}} \goodgap
\subfigure{\includegraphics[width=7cm]{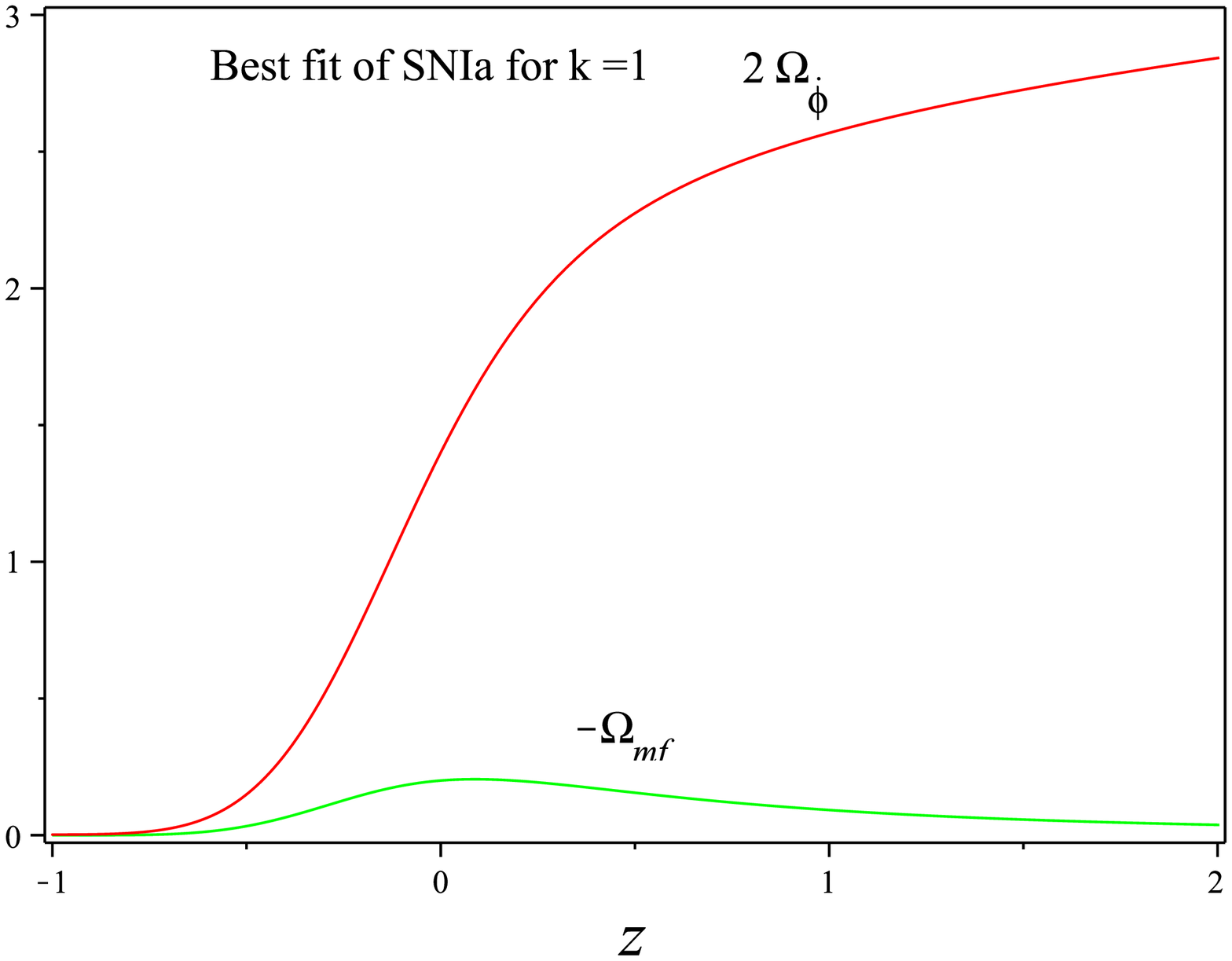}} \goodgap
\subfigure{\includegraphics[width=7cm]{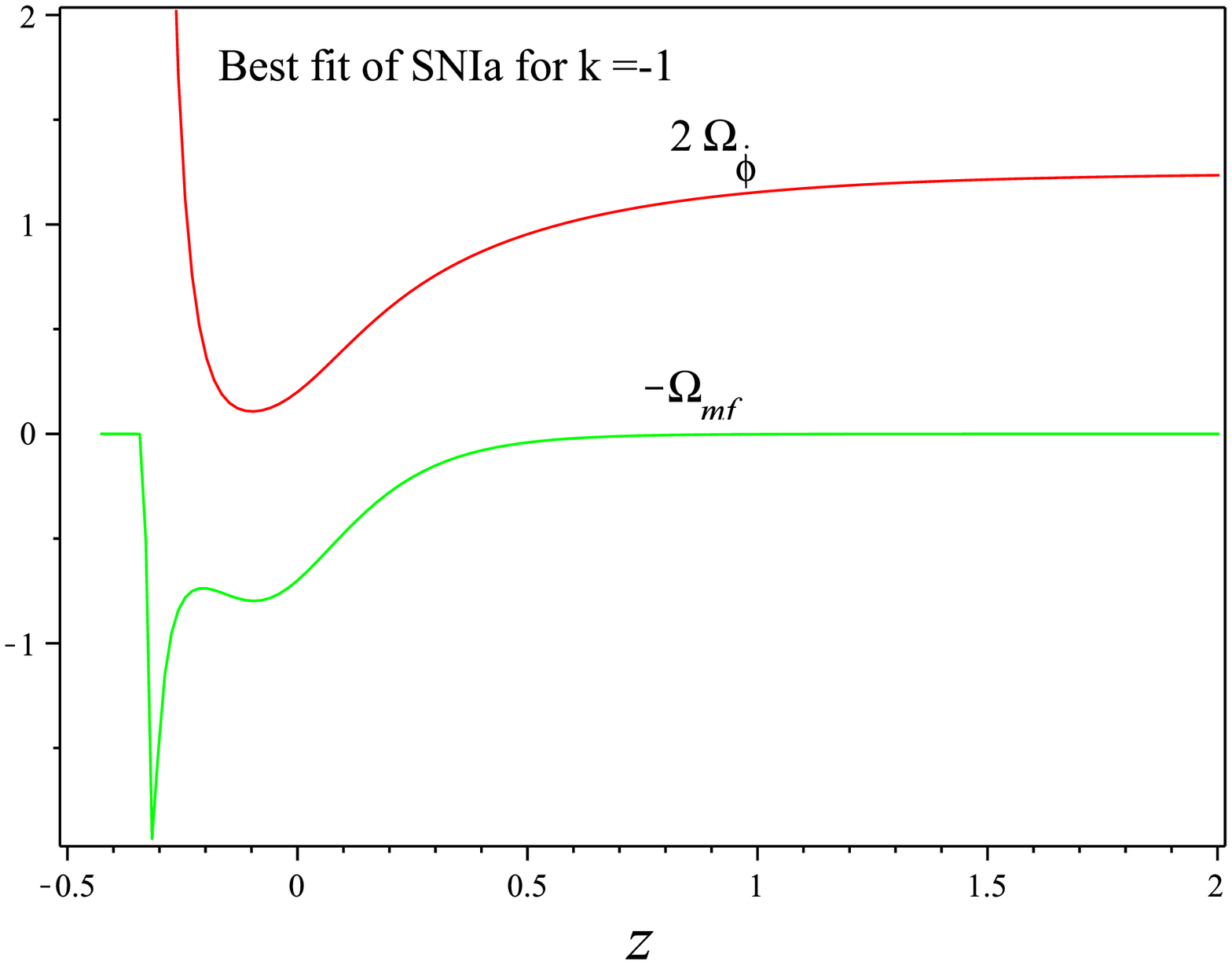}} \goodgap\\
\caption{The best-fitted effective $-\Omega_{mf}$ and, $2\Omega_{\dot{\phi}}$  plotted as function of redshift in $k=0, \pm 1$ cases }
\label{fig: clplots}
\end{figure*}

\begin{figure*}
\centering
\subfigure{\includegraphics[width=7cm]{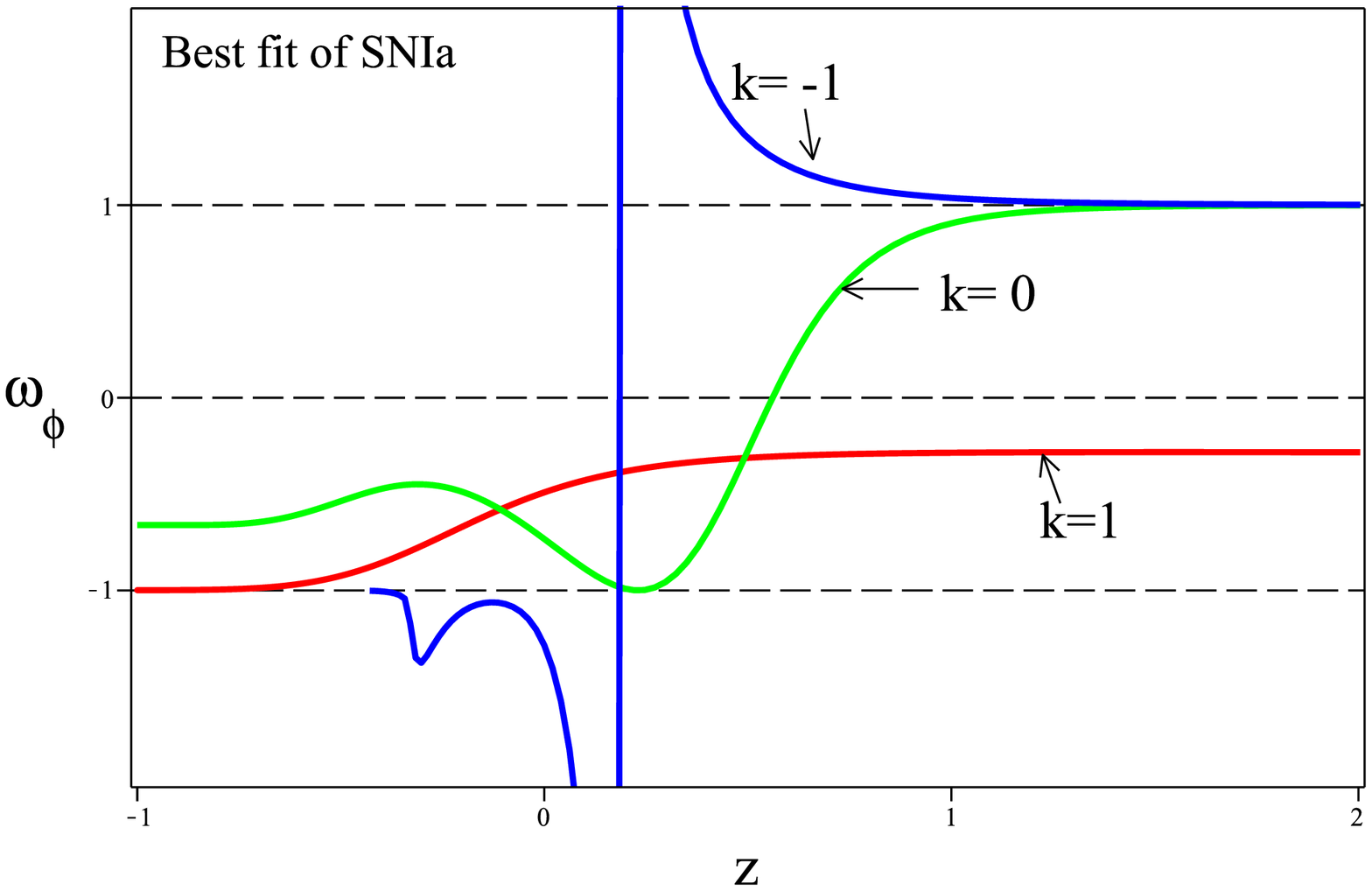}} \goodgap
\subfigure{\includegraphics[width=7cm]{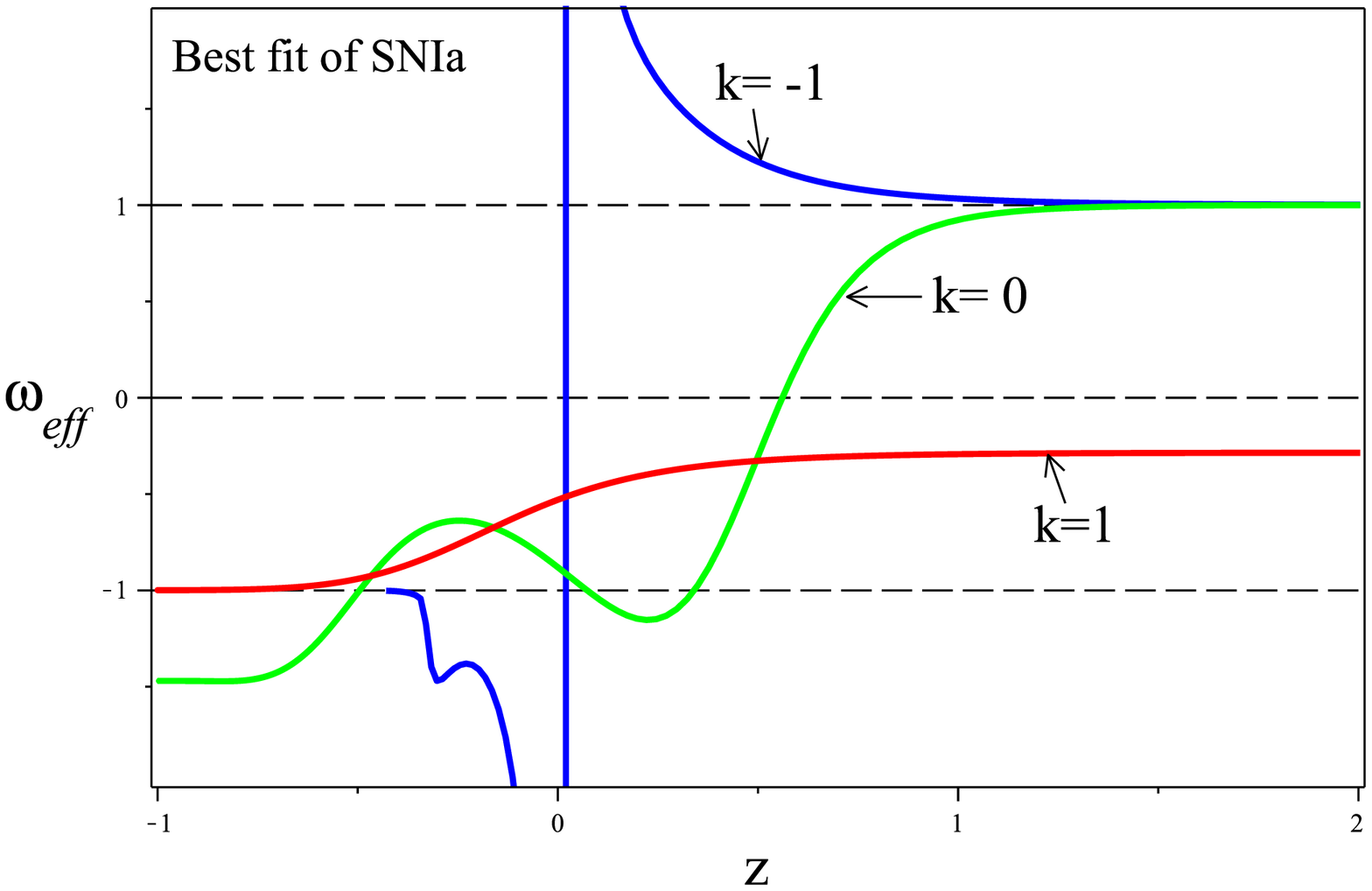}} \goodgap
\subfigure{\includegraphics[width=7cm]{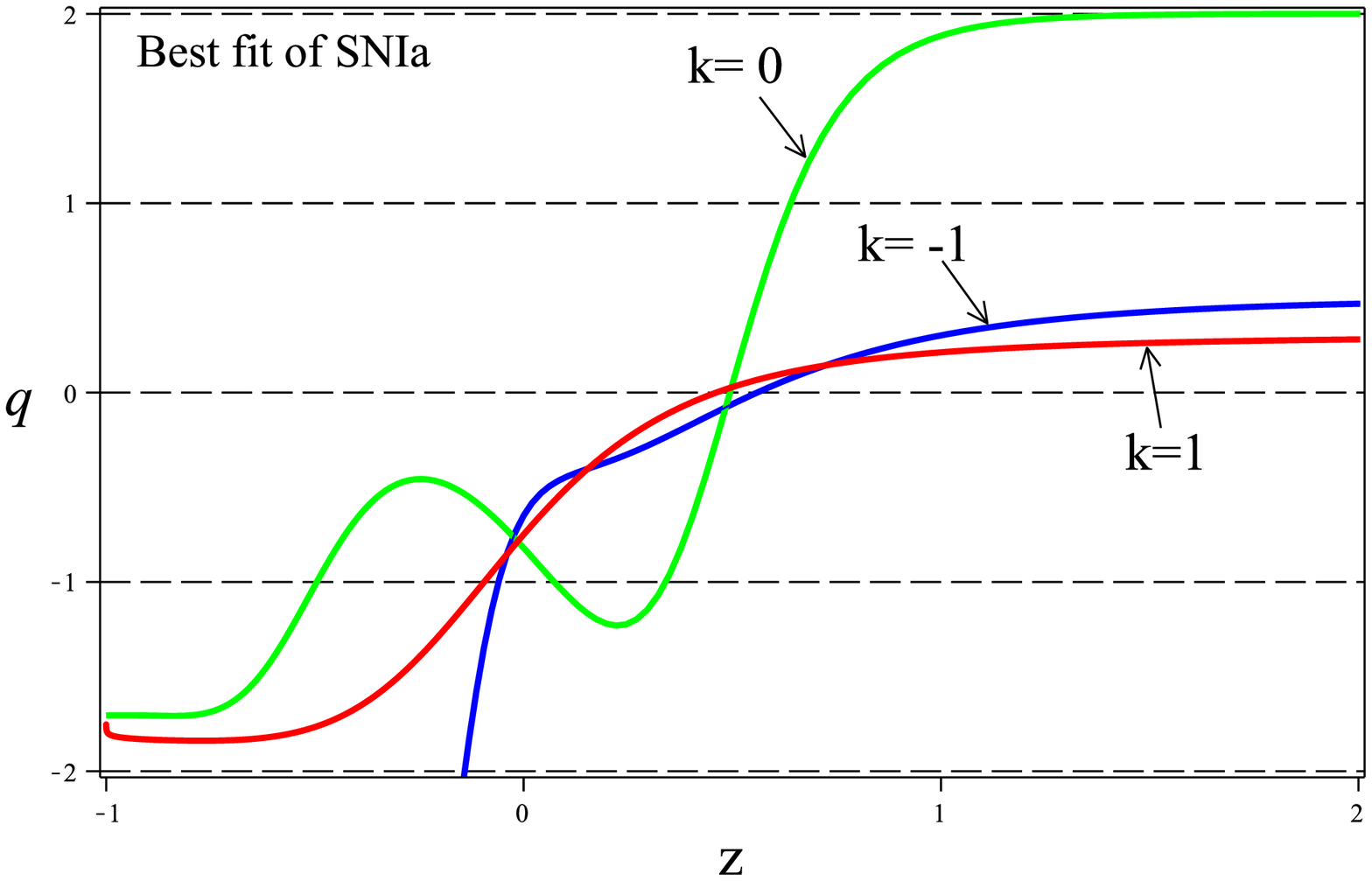}} \goodgap\\
\caption{The best-fitted effective EoS parameter, scalar field EoS parameter and deceleration parameter plotted as function of redshift in $k=0, \pm 1$ cases }
\label{fig: clplots}
\end{figure*}

The dynamics of the interacting term, $Q$, and the ratio $r$ are plotted in Fig. 7, for $k=0, \pm 1$ cases. The sign of $Q$ for different cosmological models depends on the sign of $f(\phi)$ and $\dot{\phi}$ (Fig. 8). According to the plots in Fig.8, one finds that $Q\lessgtr 0$ for $k=\pm 1$, since $\dot{\phi}$ is positive in both cases and $f\lessgtr 0$ respectively. It means that for $k=-1$ energy transfers from coupling $\rho_m f$ field to the scalar field fluid and visa versa for $k=+1$. In flat universe, sign of the interaction term $Q$ changes from negative in the past at $z\simeq 0.25$ to positive in the near past and future, as $f(\phi)$ is always negative and $\dot{\phi}$ changes its sign at the same redshift. One may consider that the coupling $\rho_m f$  and scalar field act like dark energy and  dark matter respectively. The ratio between densities for flat and closed model of the universe are negative whereas for the open universe is positive (Fig. 7). This is due to the presence of the function $f(\phi)$ in the relation that as can be seen in Fig. 8), for open universe is positive and for flat and closed universe is always negative.\\

\begin{figure*}
\centering
\subfigure{\includegraphics[width=7cm]{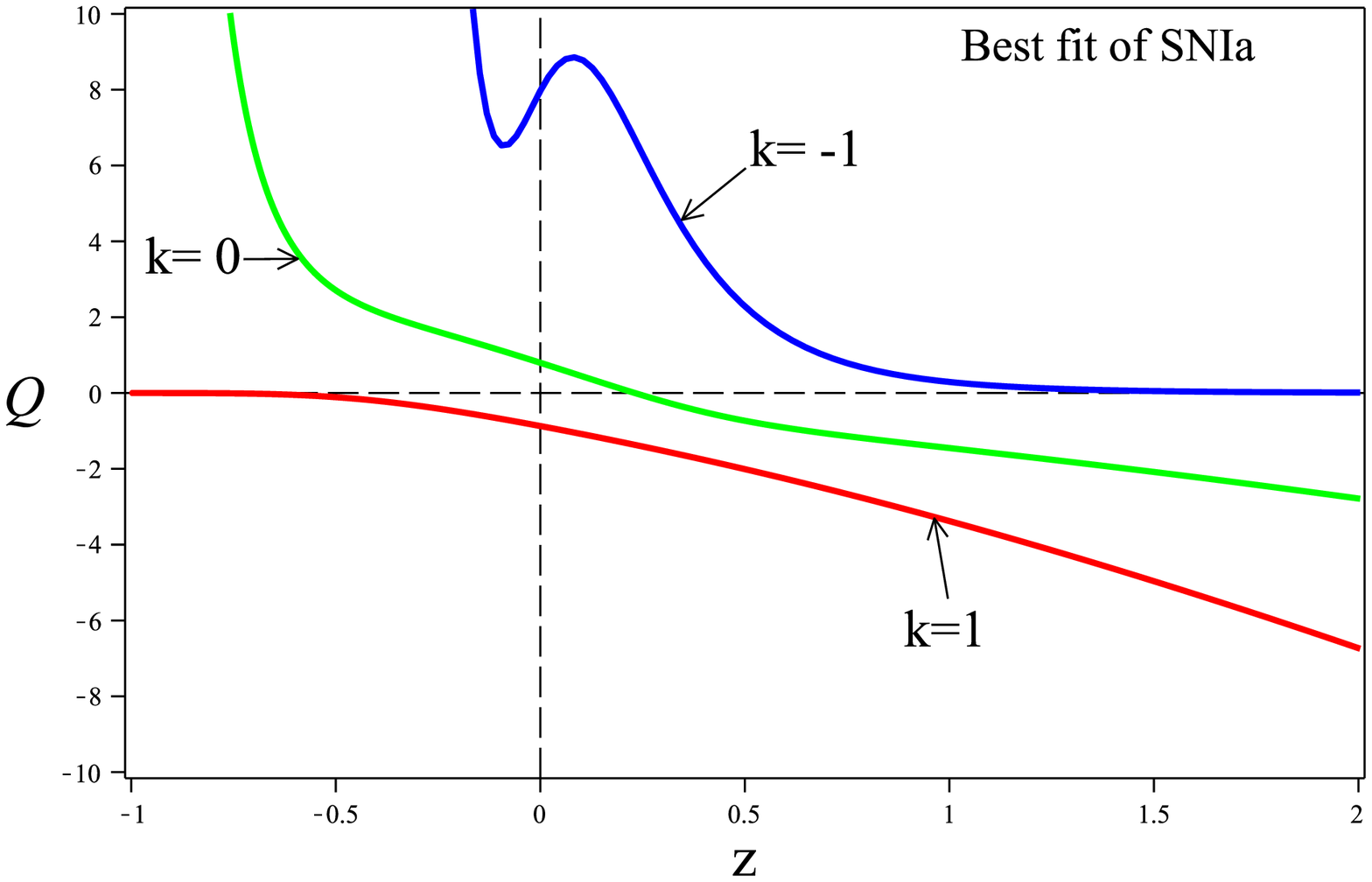}} \goodgap
\subfigure{\includegraphics[width=7cm]{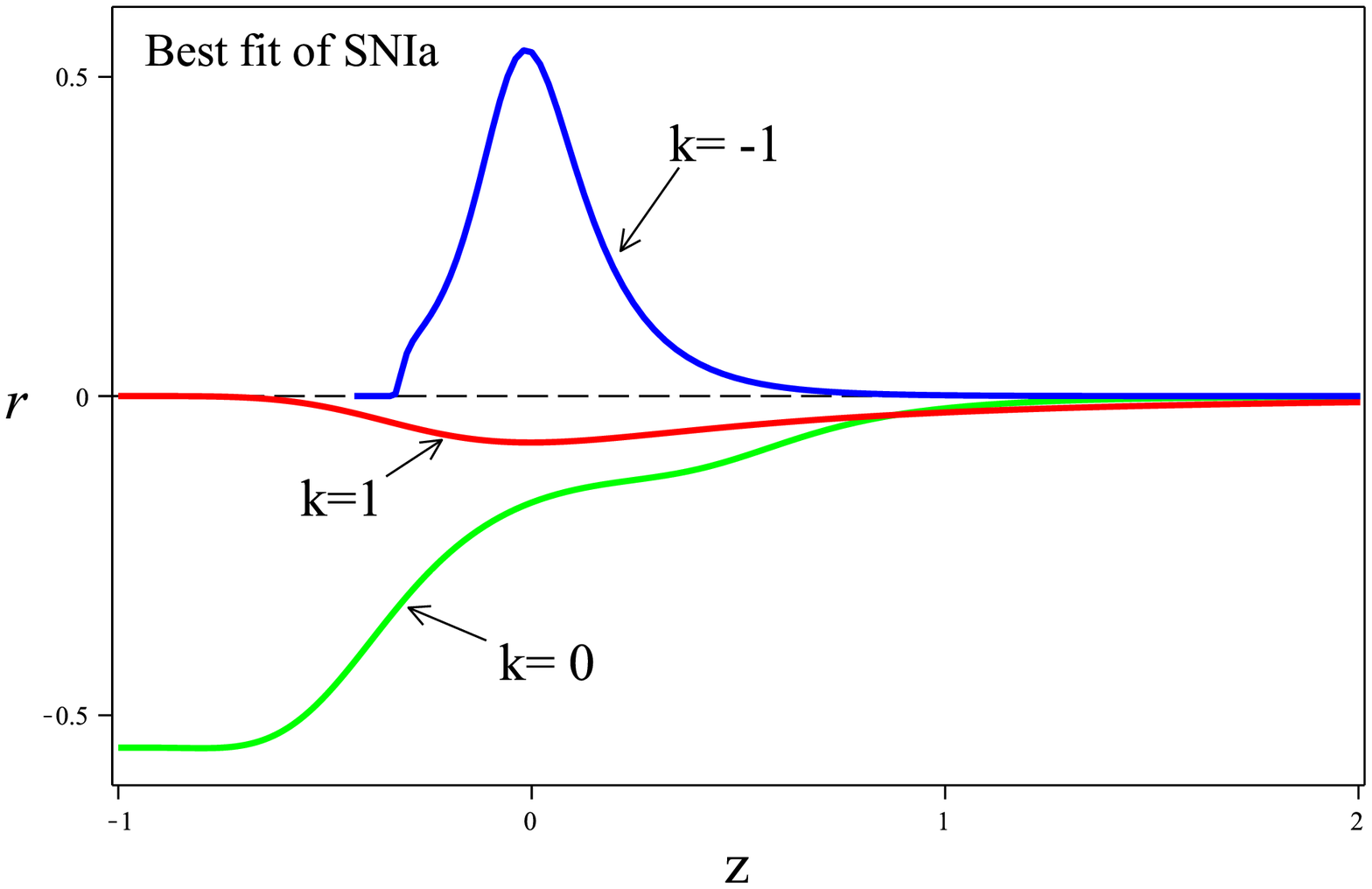}} \goodgap
\caption{The best-fitted interaction term $Q$ and densities ratio $r$ plotted as function of redshift in $k=0, \pm 1$ cases }
\label{fig: clplots}
\end{figure*}

The evolutionary form of the best fitted coupling $\rho_m f$, the reconstructed derivative of the scalar field $\dot{\phi}$ and potential $V(\phi)$ are plotted in Fig. 8, for different cosmological models. The
reconstructed scalar potential for open universe in Fig. 8 while vanishes in far past, become negative in the late time and tends to minus infinity in the near future. In case of open and flat universes, the potential goes to $+\infty$ in the near future.

\begin{figure*}
\centering
\subfigure{\includegraphics[width=7cm]{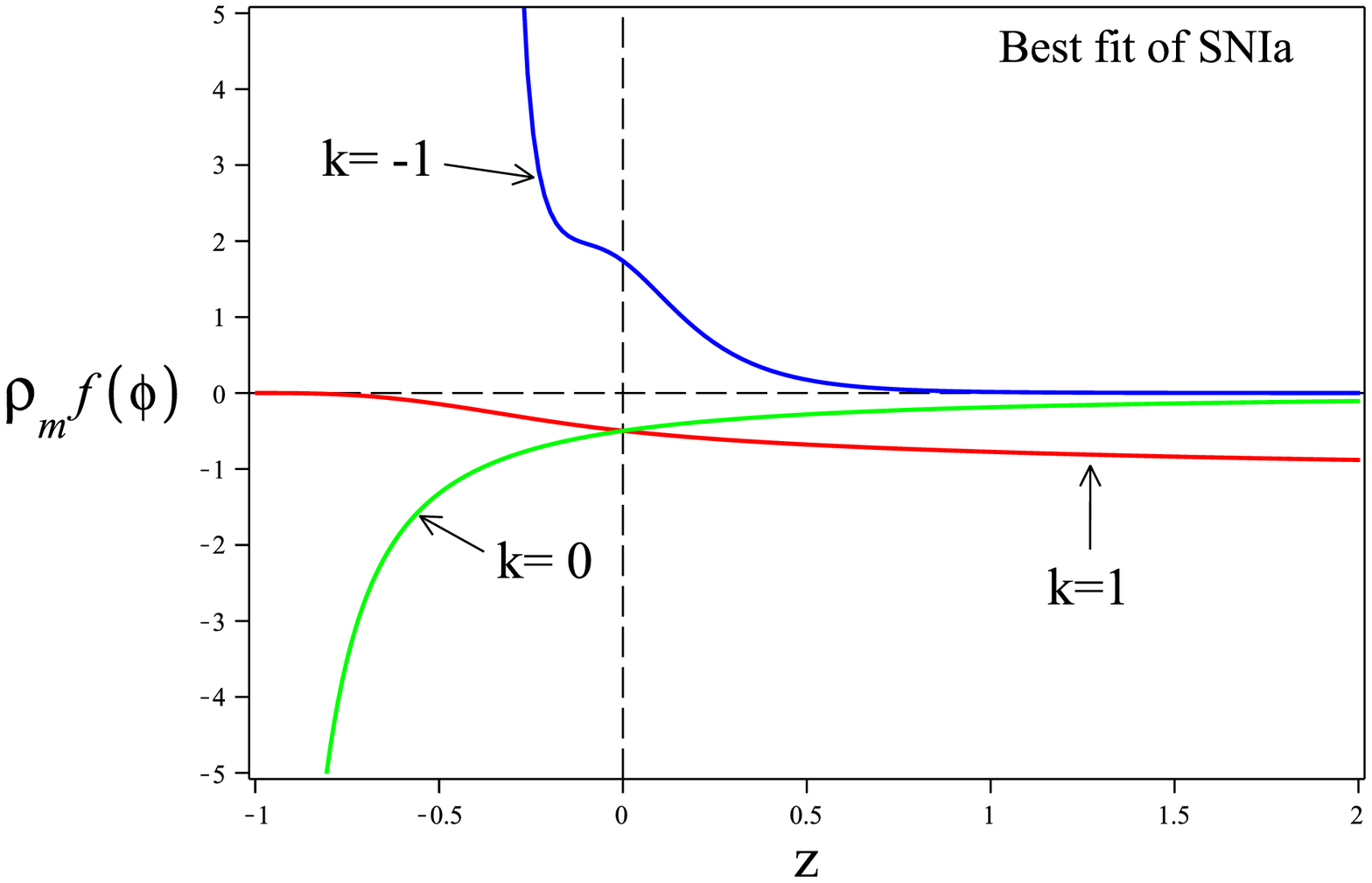}} \goodgap
\subfigure{\includegraphics[width=7cm]{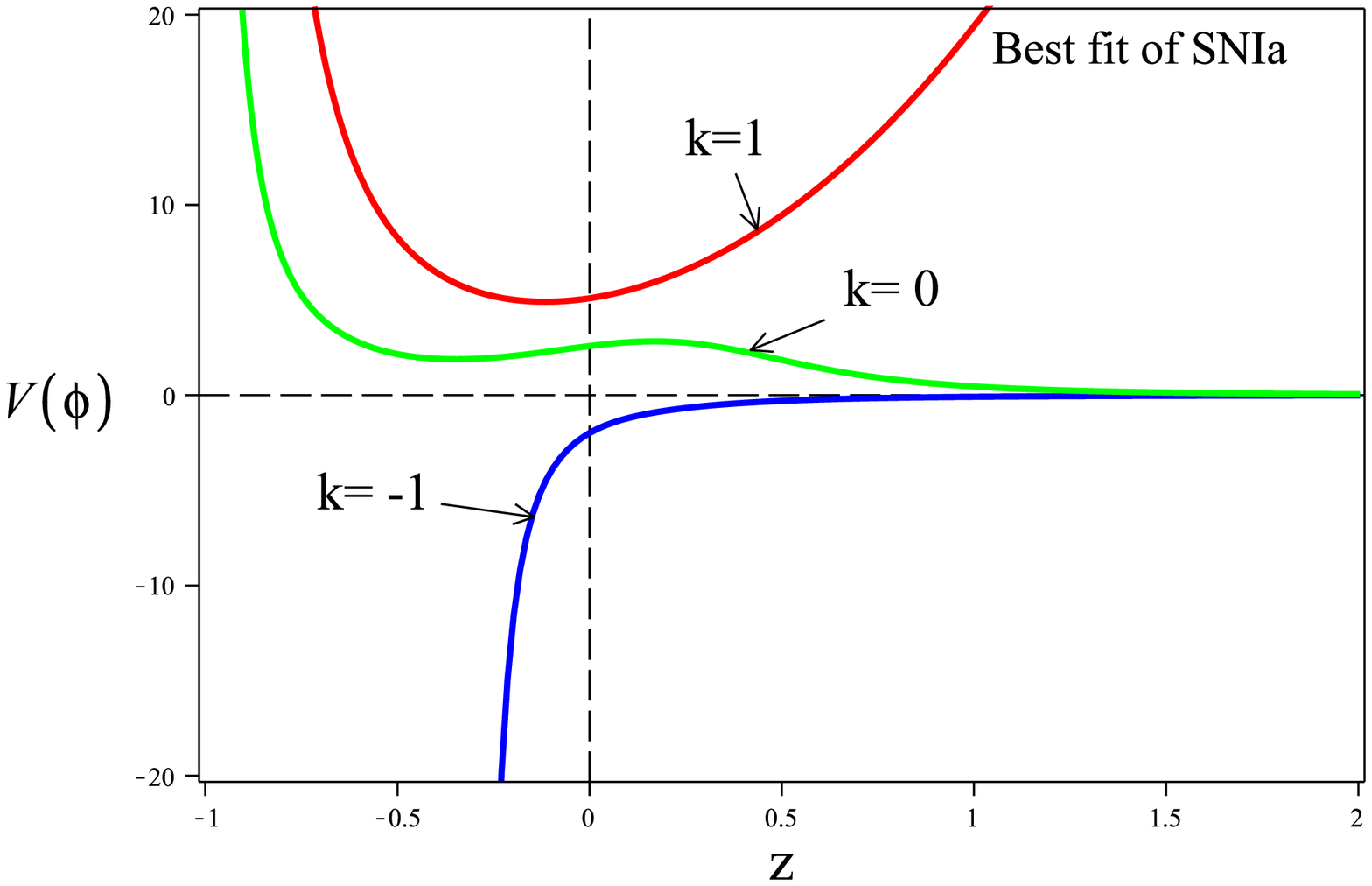}} \goodgap
\subfigure{\includegraphics[width=7cm]{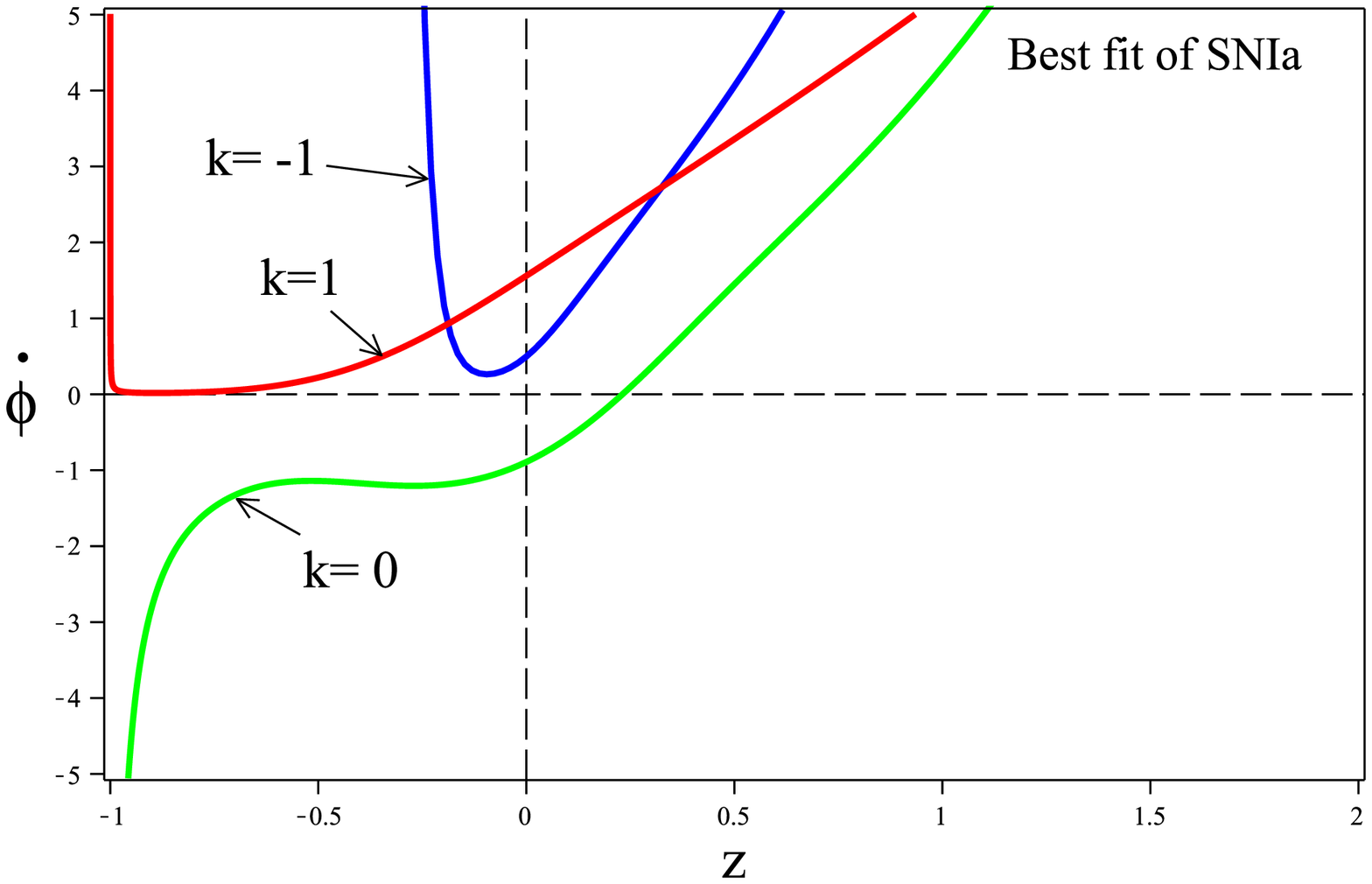}} \goodgap\\
\caption{The best-fitted of $\rho_m f$, $V(\phi)$ and $\dot{\phi}$ plotted as function of redshift in $k=0, \pm 1$ cases }
\label{fig: clplots}
\end{figure*}\

In general, the interaction term $Q$ can be an arbitrary function of cosmological parameters
like hubble parameter and energy densities. In here, we assume three forms of $Q$ as
\begin{eqnarray}\label{sigma}
Q=\sigma_{mf}H\rho_{m}f,\ \ \ Q=\sigma_{\phi}H\rho_{\phi},\ \ \ Q=\sigma_{eff}H\rho_{eff},
\end{eqnarray}
where the interaction term is proportional to the three kinds of energy densities introduced in our model. From the best fitted interaction term derived in the model and with the observational data, one can describe the dynamics of the parameters, $\sigma_{mf}$, $\sigma_{\phi}$ and $\sigma_{eff}$. In Fig. 9, the parameters are plotted against redshift $z$ for the cosmological scenarios $k=0, \pm 1$. One common feature for all these parameters is that they are flattening in high redshifts. In closed universe scenario, the parameters vanish in future. The parameters are flat in the past and only begin evolving in the near past.

\begin{figure*}
\centering
\subfigure{\includegraphics[width=7cm]{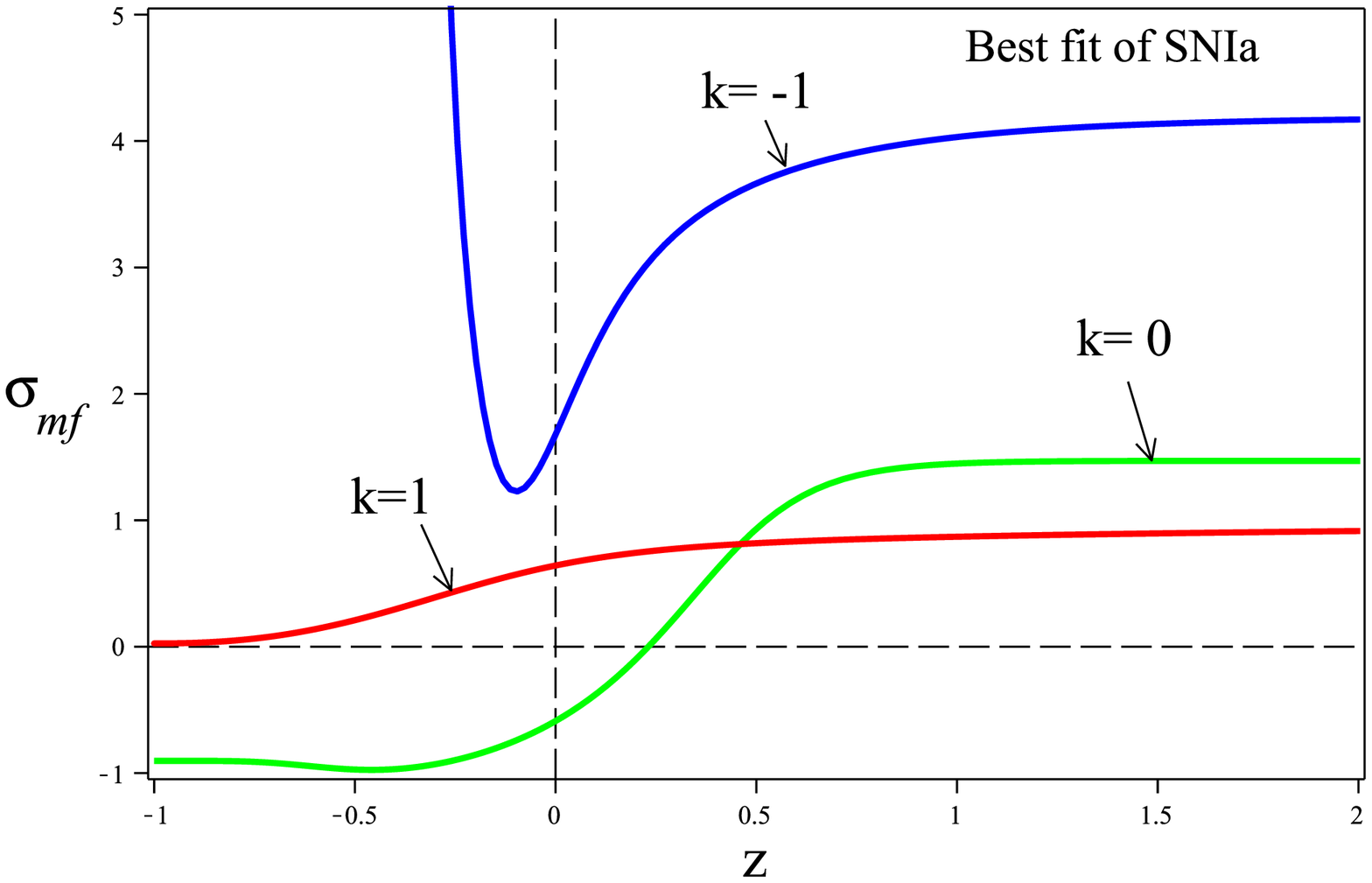}} \goodgap
\subfigure{\includegraphics[width=7cm]{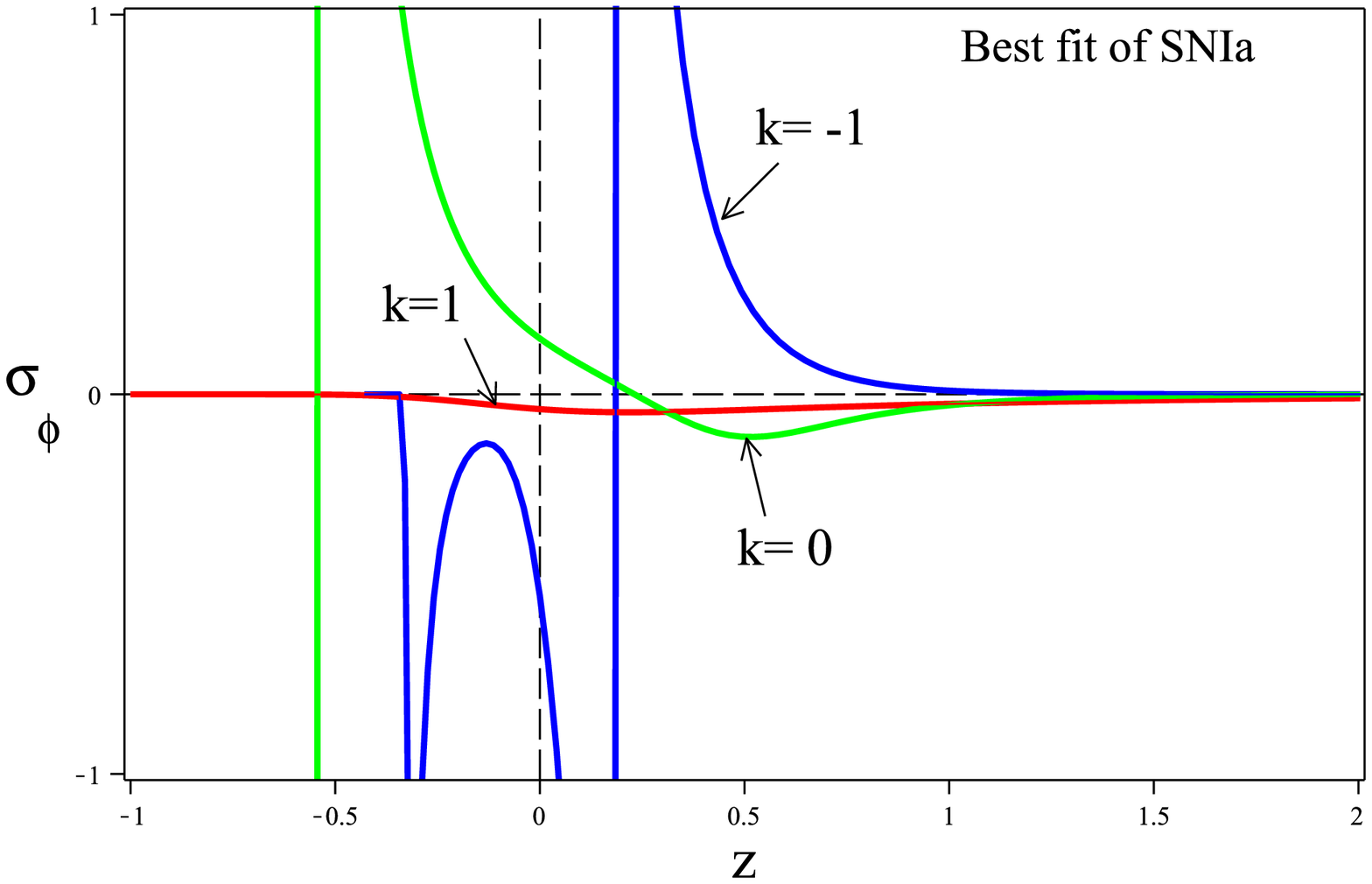}} \goodgap
\subfigure{\includegraphics[width=7cm]{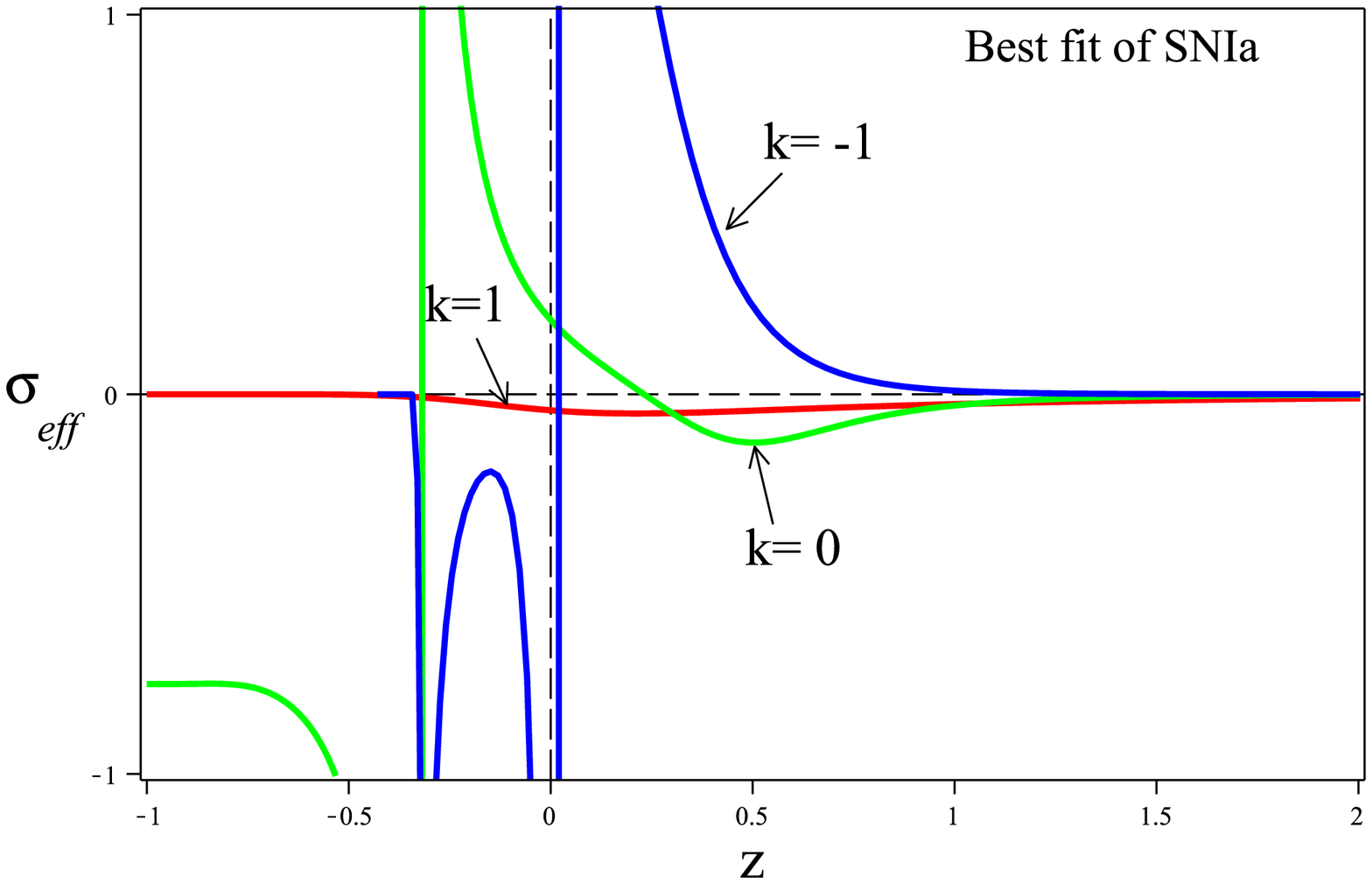}} \goodgap\\
\caption{The best-fitted $\sigma_{mf}$, $\sigma_{\phi}$ and $\sigma_{eff}$ parameters plotted as function of redshift in $k=0, \pm 1$ cases }
\label{fig: clplots}
\end{figure*}

\section{conclusion}

In this paper, we have studied the evolution of the gravitational and scalar field in cosmology in terms of dimensionless dynamical variables, in which a scalar field is
nonminimally coupled to the matter lagrangian. The field equations are solved simultaneously by means of best fitting the model parameters with the data obtained from the observations of type
Ia supernovae and utilizing the $\chi^2$ method for the distance modulus. This approach imposes observational constraints on the model parameters and yields physically more accurate solutions and observationally more reliable results. We present the model for the cases of flat, closed and open types of the universe.

The best fitted model parameters in flat universe for the effective EoS parameter show twice phantom crossing in the past and once in the future, whereas no crossing occurs for open and closed models of the universe. There are two necessary conditions need to be satisfied for phantom crossing, i.e. 1) $f(\phi)<0$, 2) $2\Omega_{\dot{\phi}}=-\Omega_{mf}$, that only in flat universe can be fulfilled.  The
graph for evolution of the effective EoS parameter in case of flat universe reveals a multiple transition from $\omega_{eff}>-1$ (quintessence era) in the
past to $\omega_{eff}<-1$ (phantom era) in the recent past, again to $\omega_{eff}>-1$ very near past and finally to phantom era in future. The phantom crossing for $k=0$ occurs within the range of the observationally accepted redshift $z < 0.45$. The best fitted deceleration parameter shows that in case of open universe the model predicts big rip in near future.

\end{document}